\documentclass[longauth, utf8]{aa} 
\usepackage[utf8]{inputenc}
\usepackage{multirow}
\usepackage{gensymb}
\usepackage{placeins}

\usepackage{graphicx}
\usepackage{txfonts,textcomp}
\DeclareUnicodeCharacter{0308}{\"{}}
\usepackage[colorlinks=true,citecolor=blue]{hyperref}

\newcommand{\vspacetab}{\rule{0pt}{3ex}}

\begin{document}

   \title{VLTI/MATISSE observations of the hot dust around \object{$\beta$ Pictoris}\thanks{Based on observations collected at the European Southern Observatory under ESO programmes 106.21Q8.006, 106.21Q8.004, 106.21Q8.001, 110.246D.003.}}
   \subtitle{Observing faint objects with MATISSE}

\author{ 
P.~Priolet\inst{1} \and 
J.-C.~Augereau\inst{1} \and 
J.~Milli\inst{1} \and 
A.~Matter\inst{2} \and 
B.~Lopez\inst{2} \and 
H.~Beust\inst{1} \and 
J.~Varga\inst{3,4} \and 
P.~Boley\inst{5} \and 
W.-C.~Danchi\inst{6} \and 
L.~Foteini\inst{3,4} \and 
T.~Henning\inst{5} \and 
M.~Houllé\inst{1,2} \and 
M.~Letessier\inst{1} \and 
{F.~Millour} \inst{2} \and
J.~Scigliuto\inst{2} \and 
G.~Weigelt\inst{7} \and 
S.~Wolf\inst{8} \and 
O.~Absil\inst{9} \and 
J.-P.~Berger\inst{1} \and 
Y.-I.~Bouarour\inst{1} \and 
G.~Bourdarot\inst{10} \and 
D.~Defrère\inst{11} \and 
C.~Desgrange\inst{12} \and 
J.-B.~Le~Bouquin\inst{1} \and 
{D.~Iglesias} \inst{13,14} \and
{T.~A.~Stuber} \inst{16} \and
P.~Berio\inst{2} \and 
F.~Bettonvil\inst{15} \and 
P.~Cruzalèbes\inst{2} \and 
M.~Heininger\inst{7} \and 
J.~W.~Isbell\inst{16} \and 
S.~Lagarde\inst{2} \and 
A.~Meilland\inst{2} \and 
R.~Petrov\inst{2} \and 
S.~Robbe-Dubois\inst{2}
\and { MATISSE Collaboration}
\and {NAOMI Collaboration}
} 

\institute{
Univ. Grenoble Alpes, CNRS, IPAG, F-38000 Grenoble, France\\
    \email{philippe.priolet@univ-grenoble-alpes.fr} \and
Univ. Côte d’Azur, Observatoire de la Côte d’Azur, CNRS, Laboratoire Lagrange, Nice, France \and
Konkoly Observatory, Research Centre for Astronomy and Earth Sciences, HUN-REN, Konkoly-Thege Miklós út 15-17, 1121 Budapest, Hungary \and
CSFK, MTA Centre of Excellence, Konkoly-Thege Miklós út 15-17, H-1121 Budapest, Hungary \and
Max Planck Institute for Astronomy, K"onigstuhl 17, D-69117 Heidelberg, Germany \and
NASA Goddard Space Flight Center, Astrophysics Division, Greenbelt, MD 20771, USA \and
Max-Planck-Institut f"ur Radioastronomie, Auf dem H"ugel 69, 53121, Bonn, Germany \and
Institute of Theoretical Physics and Astrophysics, University of Kiel, Leibnizstr. 15, 24118 Kiel, Germany \and
STAR Institute, University of Liège, Liège, Belgium \and
Max-Planck-Institut f"ur extraterrestrische Physik, Gießenbachstraße 1, 85748 Garching bei ̈M"unchen, German \and
Institute of Astronomy, KU Leuven, Celestijnenlaan 200D, 3001, Leuven, Belgium \and
European Southern Observatory, Alonso de Córdova 3107, Vitacura, Santiago, Chile \and
School of Physics and Astronomy, University of Leeds, Sir William Henry Bragg Building, Leeds, LS2 9JT, UK \and
Astrophysics Research Cluster, School of Mathematical and Physical Sciences, The University of Sheffield, Hounsfield Road, Sheffield, S3 7RH, UK \and
SRON Netherlands Institute for Space Research, Niels Bohrweg 4, 2333 CA Leiden, The Netherlands \and
Steward Observatory, The University of Arizona, 933 North Cherry Avenue, Tucson, AZ 85721, USA
}
   \date{Received 4 August 2025 ; accepted 20 May 2026}

  \abstract
   {Hot exozodiacal dust has been detected in approximately 20\% of nearby main-sequence stars, indicating that it is a common feature in extrasolar planetary systems. However, their spatial distribution and grain properties remain poorly understood.}
   {We  aim to show that the spatial distribution of hot dust around $\beta$~Pictoris\ -- an iconic planetary system with two known planets and a population of exocomets -- can be effectively constrained through a detailed analysis of optical long-baseline interferometry data.}
   {We obtained \textit{L}-band ($\lambda \sim 3.5~\mu$m) observations of $\beta$~Pictoris\ with the MATISSE instrument at the VLTI. We first developed a method to assess correlations in the visibility measurements by estimating the full data covariance matrix. We then modeled the circumstellar emission using geometric brightness distributions to fit the observed visibilities.}
   {This study reveals circumstellar emission around $\beta$~Pictoris\ consistent with hot exozodiacal dust. By modeling the emission both accounting and not accounting for correlations in the data, we demonstrate their significant impact on the inferred dust properties. The best-fit model reveals a compact ($\sim$0.05 au), highly inclined dust structure near the sublimation radius, misaligned by $\sim 60\degree$ on the sky plane with respect to the extended edge-on outer disk, contributing $\sim 4\%$–$7.5\%$ of the total flux at $\lambda = 3.4~\mu$m, and is composed of submicron-sized grains. The data { are also compatible with an additional} dust component near 1~au.}
   {In the case of $\beta$~Pictoris, the two-decade-old paradigm -- placing hot exozodiacal dust near the sublimation radius and composed of submicron-sized grains -- finds observational support in the detailed analysis of MATISSE data. The misalignment with the known debris disk further supports a link to the exocometary activity.}


   \maketitle

\section{Introduction}
Debris disks are systems composed of small rocky bodies such as comets, asteroids, dwarf planets, and dust orbiting a main-sequence star. In the Solar System, these components include the Kuiper Belt, the asteroid belt, as well as the zodiacal cloud and F-corona near the Sun. In extrasolar systems, distant cold dust produced by collisions between small rocky bodies was first identified through the detection of far-infrared (FIR) excess emission around Vega \citep{firstdetectdebrisdisk}. Since this discovery, debris disks have been found around approximately 20\% of main-sequence stars \citep{debrisdiskstats}. These disks typically manifest as belts located tens to hundreds of au from their central star \citep[e.g.,][]{Matra2025}. Some also contain cold gas, which is hypothesized to be released by volatile-rich, comet-like objects \citep[e.g.,][]{Kral2020}.

However, the characterization of the dust and gas components in the inner regions of extrasolar planetary systems remains challenging. In this context, the first detection of hot exozodiacal dust (hereafter exozodi) around Vega using optical interferometry represented significant advancement \citep{Absil2006}. Subsequent near-infrared interferometric surveys revealed that 15–20\% of nearby main-sequence stars exhibit excess emission due to hot dust at approximately 1\% above the stellar photosphere level in the \textit{K} and \textit{H} bands \citep[CHARA/FLUOR and VLTI/PIONIER, respectively; ][]{Absil2013, Ertel2014, Nunez2017, Absil2021}. Current models suggest that these near-infrared excesses are produced by populations of submicron-sized, likely carbonaceous, dust grains located near their sublimation zone, typically a few stellar radii from the host star \citep[e.g.,][]{Absil2006, Difolco2007, Defrere2011, Lebreton2013, Kirchschlager2017}. The persistence of such large amounts of hot dust in extrasolar systems is enigmatic. At these distances, the dust is expected to sublimate quickly, be expelled by radiation pressure, or erode through collisions \citep[e.g.,][]{Kobayashi2009, Lebreton2013, Vanlieshout2014}. This has led to the hypothesis that the hot dust is replenished by star-grazing exocomets originating from an outer planetesimal belt \citep{Marboeuf2016, Sezestre2019, Pearce2022}, {although significant theoretical barriers have been identified for this model \citep{Pearce2022}}. Furthermore, these conclusions are based on limited observational evidence \citep[see][for a recent review]{ertel2025}. In particular, current data do not provide direct constraints on the spatial distribution of the hot dust.
{

Recent MATISSE observations have studied other hot-dust systems. $\kappa$ Tuc was observed by \cite{Kirchschlager2020}, while Fomalhaut was observed by \cite{Ollmann2025}, with both studies reporting compact hot dust emission consistent with dust located near the sublimation region. These studies highlight MATISSE's potential to detect and characterize hot exozodi in nearby systems.
}

In this study, we present \textit{L}-band VLTI/MATISSE observations of $\beta$~Pictoris, aiming to characterize the geometry of its faint hot dust emission for the first time. $\beta$~Pictoris\ is a key target for studying planetary systems (see Table~\ref{tab:bpicproperties} for its basic properties). It was the first system where a cold debris disk was imaged \citep{1984betpicimg} and hosts two known exoplanets: $\beta$~Pictoris\ b, with a semimajor axis of approximately 10~au \citep{Betapic_b_paper1,Betapic_b_paper2}, and $\beta$~Pictoris\ c, with a semimajor axis of around 2.7~au \citep{Betapic_c}. It is a young ($\sim 20$~Myr), A6V-type main-sequence star, also known for its well-documented exocometary activity \citep[e.g.,][]{Ferlet1987, Kiefer2014, Lecavelier2022}. \citet{Defrere_2012_betapic} detected an excess of circumstellar emission in the \textit{H} band using VLTI/PIONIER in 2010 and 2011, amounting to $1.37 \pm 0.16\%$ of the stellar photospheric flux. A similar level of excess was consistently observed over the following three years with the same instrument \citep{Ertel2016}. This emission has been attributed to hot dust located in the inner regions of the debris disk, within the orbits of known exoplanets (<3 au).

The VLTI/MATISSE observations of $\beta$~Pictoris\ presented in Sect.~\ref{sec:observations} reveal a faint excess emission in the \textit{L} band, which we attribute to hot exozodi. In Sect.~\ref{sec:correlations}, we introduce an original approach to estimate correlations in the MATISSE \textit{L}-band data, enabling careful modeling of the observations. We then explore in Sect.~\ref{sec:Models} four geometric models for the exozodiacal brightness distribution, demonstrating that the MATISSE data provide valuable insights into the geometry of the detected excess. The implications of these results are discussed in Sects.~\ref{sec:discussion} and \ref{sec:discussion_interfero}.


\begin{figure*}[tbp!]
  \centering
  \hbox to \textwidth {
  \parbox{0.63\textwidth}{\includegraphics[width=0.63\textwidth,origin=bl]{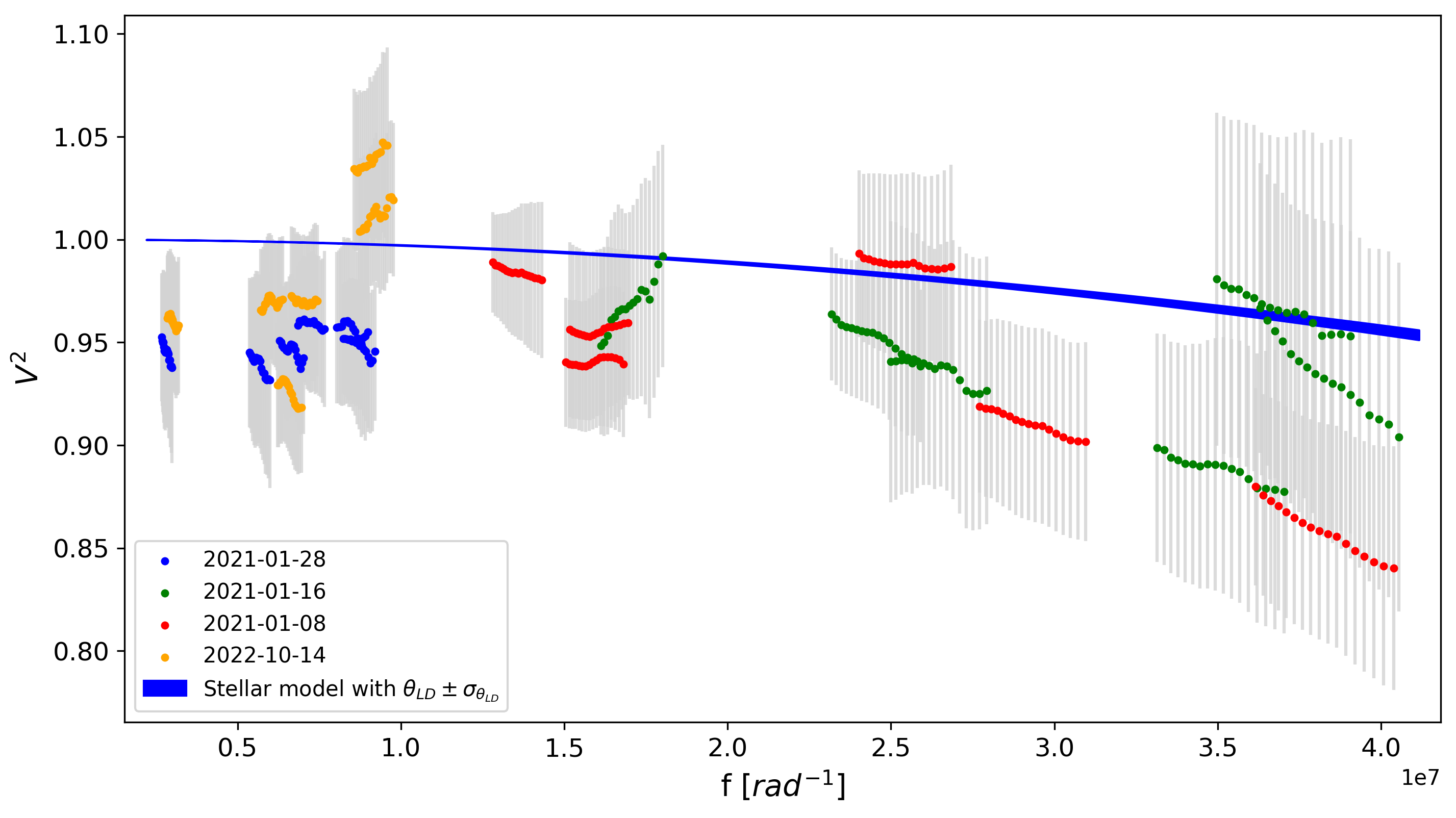}}
  \parbox{0.37\textwidth}{\includegraphics[width=0.37\textwidth,origin=bl]{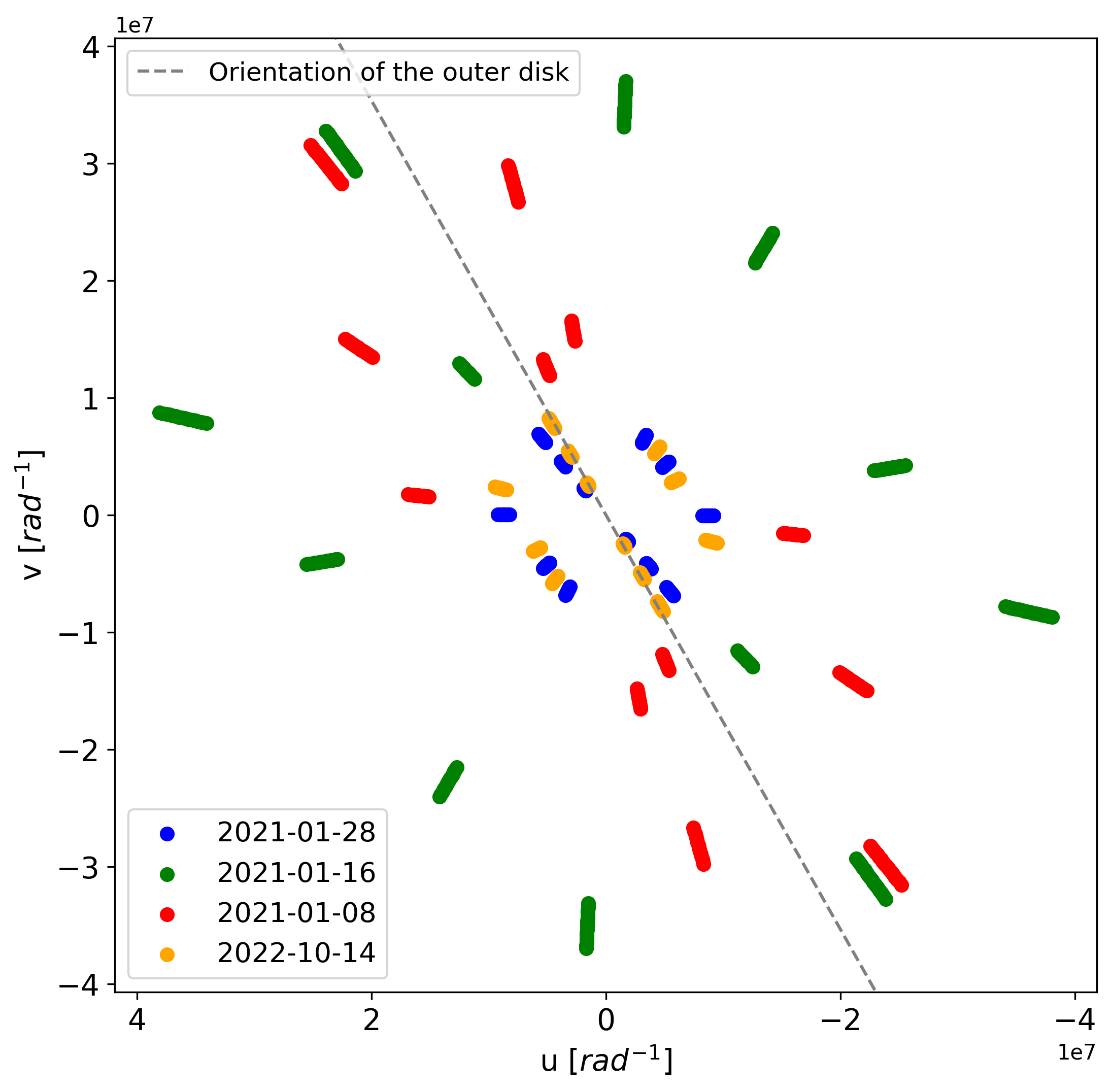}}
  } 
  \caption{\textsl{Left panel:} \textit{L}-band interferometric squared fringe visibilities ($V^2$) as a function of spatial frequency ($B/\lambda$, where $B$ is the projected baseline) for all the MATISSE observations in Table~\ref{tab:observations}. The blue-shaded area represents stellar squared visibility for the $\beta$~Pictoris\ stellar model with a 1$\sigma$ uncertainty on the angular diameter $\theta_{\rm LD}$ (see Table~\ref{tab:bpicproperties}). \textsl{Right panel:} $uv$ plane coverage of all the observations. The dashed line represents the orientation of the major axis of the outer disk on the sky (PA=$29.6 \degree$;
 see Table~\ref{tab:bpicproperties}).
  \label{fig:dataanduvplane}}
\end{figure*}

\section{Observations}
\label{sec:observations}

Interferometric observations of $\beta$~Pictoris\ were obtained in January 2021 and October 2022 using the Multi-AperTure mid-Infrared SpectroScopic Experiment \citep[MATISSE,][]{MATISSE} at the Very Large Telescope Interferometer (VLTI) on Cerro Paranal, Chile. The VLTI can combine the light from either the four 8.2-m Unit Telescopes (UTs) or the four movable 1.8-m Auxiliary Telescopes (ATs), providing different baseline lengths and angular resolutions depending on the array configuration. These observations, detailed in Table~\ref{tab:observations}, were conducted in low spectral resolution ($\lambda/\Delta \lambda \sim 30$) using the GRAVITY fringe tracker (GRA4MAT)  and obtained across a wide range of telescope configurations, encompassing both short and long baselines in the \textit{L}, \textit{M}, and \textit{N} bands (3.2--3.9~$\mu$m, 4.5--5.0~$\mu$m, 8--13~$\mu$m, respectively). They were interleaved with observations of reference stars to calibrate the instrumental and atmospheric effects in the measured quantities. These calibrators were selected based on their proximity to $\beta$~Pictoris\ both in terms of position and magnitude, and on their angular size (which should be as small as possible). An observing block (OB) corresponds to the observation of one target. Typically two to three OBs are necessary (one science target plus one or two calibrators), with each OB lasting $\sim 30$ min. Details on the sequence used for the observations in this study are given in Table~\ref{tab:observations}.

The data were processed using the standard EsoRex Data Reduction Software (DRS) for MATISSE \citep{std_MAT_DRS}. Here, we used only the \textit{L}-band data as they have the higher signal-to-noise ratio (S/N) required for this type of study. An observation with MATISSE can be performed in two different ways, with and without chopping. Chopping consists in alternating between the target and the sky background to better subtract the latter from the observations. In a few cases, we only used the non-chopped data because they increase fringe tracking stability and yield a better S/N.  In our case the observations were executed both with and without chopping. In the modeling we only used one or the other depending on the quality of the data (information on the mode used for each observation can be found in Table \ref{tab:observations}).

We focus here on the interferometric fringe squared visibilities ($V^2$), shown in the left panel of Fig.~\ref{fig:dataanduvplane}, across 17 spectral channels between $\lambda = 3.2~\mu$m and $3.6~\mu$m, measured for each of the six baselines in the four datasets listed in Table~\ref{tab:observations}. These data suggest a visibility deficit, which we attribute to an extended circumstellar emission (see Sect.~\ref{sec:Models}). On the other hand, the closure phases are close to zero for all the datasets used (see Sect.~\ref{sec:closure_phase}), which allowed us to exclude the presence of a bright companion in the instrument's field of view (FOV) as the source of the excess emission detected\footnote{The FOV is defined by the diameter of the MATISSE pinhole, which, in the \textit{L} band, is $\sim 600$~mas with the ATs and $\sim 130$~mas with the UTs}. Furthermore, a companion bright enough to produce the observed visibility deficit would likely have already been detected, as it would induce a measurable radial velocity signal (see \cite{ertel2025} for details). This is especially true in the case of $\beta$~Pictoris, which is viewed nearly edge-on. The closure phase (CP) also makes it possible to rule out a strong asymmetry in the circumstellar emission. The $uv$ plane of all observations is shown in the right panel of Fig.~\ref{fig:dataanduvplane}. 
\section{Covariance matrix and errors} \label{sec:correlations}

In our modeling, we adopted a parametric approach to reproduce the observed squared visibilities (see Sect.~\ref{sec:Models}). Previous studies employing this method have typically assumed that observables (such as visibility measurements) are statistically independent, implying that errors between different data points are uncorrelated. However, in optical interferometry, this assumption does not hold true. Several examples of these correlations have been demonstrated, such as for calibration errors \citep{Perrin2003}, closure phases \citep{CP_redundancy}, and the squared visibility amplitudes with VLTI/GRAVITY in the context of exoplanets \citep{GRAVITY_correlations}. 

Given that hot exozodis are at the sensitivity limit for current interferometric instruments (the circumstellar emission accounting for only a few percent of the total flux at the observing wavelength), we believe that correlations in the data may introduce biases in the parameters inferred from the disk models. In \cite{Absil2006}, correlations due to the uncertain angular size of the calibrator were taken into account in the modeling of the circumstellar emission around Vega with CHARA/FLUOR (\textit{K} band) using the formalism from \cite{Perrin2003}. Correlations were also considered in some later exozodi studies, such as \cite{Ertel2014}, which assumed that spectral channels within one baseline are fully correlated, while different baselines are treated as fully uncorrelated. However, none of these studies directly estimated the covariance matrix from the data.

In this section, we introduce a method for estimating the correlations in VLTI/MATISSE visibility measurements. They are represented by the correlation matrix $\mathbf{K}$ and the variance-covariance matrix $\mathbf{\Sigma}$, where $\Sigma_{kl} = \sigma_k \sigma_l K_{kl}$, and $\sigma_k$ and $\sigma_l$ represent the standard deviations of measurements $k$ and $l$, respectively. In Sect.~\ref{sec:Models}, we evaluate the impact of these correlations on the retrieval of model parameters.

\begin{figure*}[tbph!]
    \centering
    \includegraphics[bb = 0 0 2560 1080, clip, width=0.98\textwidth]{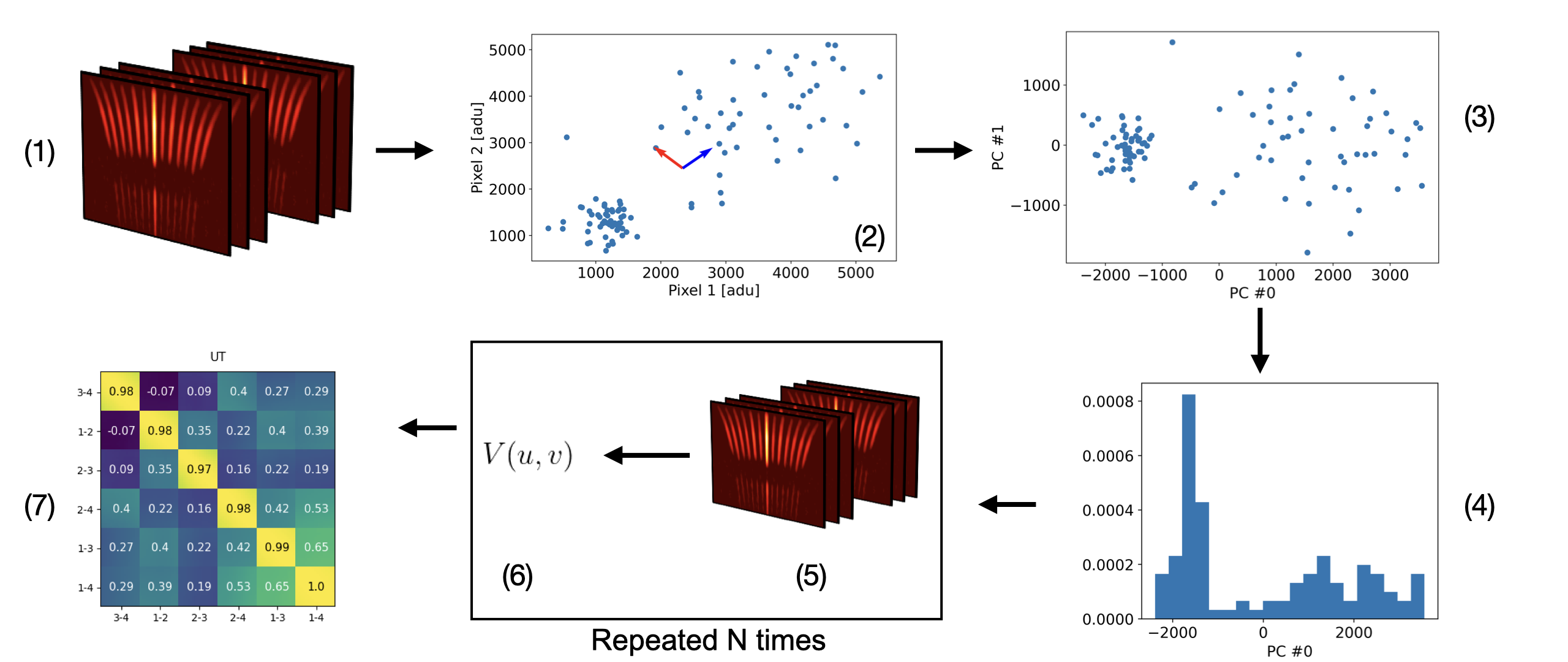}
    \caption{Diagram of the process (Sect.~\ref{sec:correlations_general_approach}) for generating visibilities used in the calculation of the correlation matrix: (1) intermediate observational frames; (2) pixel–pixel scatter plot of values with arrows showing the two PCs; (3) data of these two pixels transformed into the PC basis; (4) histogram of the values along one PC; (5) distribution used for generating synthetic frames; (6) visibilities extracted from these frames; (7) covariance matrix computed from the synthetic visibilities. Steps (5) and (6) were repeated to properly sample the visibility distribution.}
    \label{fig:recap_correlations_method}
\end{figure*}

\subsection{The final step of MATISSE data reduction}
\label{sec:final_step_of_data_reduction}
To appreciate our approach for estimating the covariance matrix, it is essential to understand the MATISSE measurement reduction process. To obtain the squared visibility using the standard MATISSE DRS, the incoherent {correlated flux} was first calculated as follows:
\begin{equation}
    C^{2}_{ij} (\lambda) = \sum_f \  \left< \left|  \underline{\textbf{I}}(B_{ij}, \lambda, t)\right|^2 -\beta \right>_t,
\end{equation}
where $C_{ij}$ is the \textit{correlated flux} for a baseline formed by telescopes $i$ and $j$, $\underline{\textbf{I}}(B_{ij}, \lambda, t)$ is the complex Fourier transform of the interferogram at wavelength $\lambda$, $B_{ij}$ is the projected baseline, $\beta$ is the bias estimated on $\textrm{I}$ outside of the fringe peaks, $\left< \right>_t$ denotes the time-average performed over the duration of the exposure ($\sim $30--60~s), and {$f$ is the spatial frequency for all pixels between $B_{ij}-D$ and $B_{ij}+D$ where $D$ is the diameter of a single telescope}. Here, an exposure refers to a collection of integrations in a particular instrumental setup \citep[see][for more information]{MATISSE}. This squared {correlated flux} $C_{ij}^2$ was then divided by a photometric term to obtain the squared visibilities $V^2_{ij}$:
\begin{equation}
    \label{eq:visibility_MATISSE}
    V_{ij}^2 (\lambda) = \frac{C_{ij}^2(\lambda)}{ \sum_x \  \left< P_{i}(x, \lambda, t)P_{j}(x, \lambda, t)\right>_t},
\end{equation}
where $P_{i}(x, \lambda, t)$ is the photometric flux for telescope $i$, corrected by the Kappa matrix\footnote{The Kappa matrix was used to correct for the difference in response between the detector pixels \citep[see][for more details]{MATISSE}}, and $x$ is the pixel index along the spatial direction of the detector. This process was repeated four times, with the beams commuted each time using a beam commuting device (BCD) {to calibrate closure and differential phases.}

The above procedure represents the final step in the data reduction cascade where the observations were handled frame by frame. During this step the frames were time-averaged, resulting in a single visibility measurement per exposure, rather than per frame. {Indeed, in the standard MATISSE DRS, the visibility is calculated as the ratio between the time average of the correlated flux and the time average of a photometric term (Eq. \ref{eq:visibility_MATISSE}). Therefore, there is no visibility measurement across time (a visibility per frame) with which to calculate the correlations as in \cite{GRAVITY_correlations}.}

Consequently, if we aim to study the correlations between data points across time, we must extract the covariance matrix of the pixels from the intermediate frames $\underline{\textbf{I}}(B_{ij}, \lambda, t)$ and $P_{i}(x, \lambda, t)$.
For each detector integration time (DIT), we obtain four photometric frames $P_{i}(x, \lambda, t)$ (one for each telescope) and two {correlated flux} frames $\underline{\textbf{I}}(u, \lambda, t)$ (the real and imaginary parts). In total there are about $\sim $100--200 frames per observing configuration (for a given BCD position for a given chopping mode).

It is instructive to observe the correlations in the intermediate frames obtained during an exposure. These are illustrated in Appendix~\ref{sec:correlations_raw_data} between wavelengths or telescopes for the photometry, as well as between wavelengths or baselines for the {correlated flux}. These intermediate frames reveal not only the very strong correlations between spectral channels within a single baseline, but also the complex nature of correlations between the correlated flux of two baselines and between the photometric measurements of two telescopes.

\subsection{General approach}
\label{sec:correlations_general_approach}
We derived the covariance matrix of the MATISSE \textit{L}-band data from the intermediate frames and propagated it to the visibilities using a Monte Carlo approach. 
Our method, illustrated in Fig.~\ref{fig:recap_correlations_method}, consists of the following steps:
\begin{enumerate}
    \item[(1)] Gathering the intermediate frames (i.e., before time-averaging) from the observations;
    \item[(2)] Extracting the temporal distribution of pixel values for each frames;
    \item[(3)] Applying a principal component analysis (PCA) to these distributions;
    \item[(4)] Deriving the probability distribution of the values for each principal component (PC);
    \item[(5)] Generating new synthetic intermediate frames that follow the same probability distribution. Steps 5 and 6 were repeated $N$ times (here, $N=200$);
    \item[(6)] Processing each synthetic frame with the MATISSE DRS to obtain visibilities;
    \item[(7)] Calculating the correlation matrix from this set of visibilities.
\end{enumerate}
By following this approach, our generated visibilities (step 6) represent different noise realizations, while following the same probability distribution of the real observations. Details on steps 3 to 7 are outlined below.
\subsection{Generating synthetic intermediate frames} \label{sec:PCA}
The challenge in generating synthetic intermediate frames (step~5) in order to calculate correlation matrices (steps~6 and 7) lies in the fact that the pixels in the intermediate frames correlate with one another, as shown in Appendix~\ref{sec:correlations_raw_data}. One approach for addressing this issue may involve computing the covariance matrix of all pixels and generating their synthetic values using a multivariate normal distribution. However, this would have involved a high-dimensional space, making a direct calculation of the covariance matrix computationally expensive. Therefore, we employed PCA to compute this more efficiently (steps~3 and 4). 

Principal component analysis, also called Karhunen-Lo\`eve transform \citep{karhunen1947, loeve1948}, is a technique in which data are linearly transformed onto a new coordinate system such that the directions of the new basis vectors (PCs) capture the largest variations in the data. This also means that the PCs are linearly uncorrelated with each other. This technique is often used as a dimensionality reduction technique, since one can remove components that explain little to no variance and therefore reduce the dimension in the data. In this study, however, we used it as a way to generate correlated samples very efficiently.

The idea is to generate the synthetic data independently along each PC and then perform the inverse transform to retrieve the original coordinate system (the pixels in the frames). Therefore, the features in our PCA correspond to the pixels in the intermediate frames and the samples correspond to the different values of these pixels with time. We applied this methodology to the modulus of the {correlated flux} (rather than its real and imaginary parts; see Appendix~\ref{sec:distribution_along_PC}) as well as to the photometry of each telescope.

We sampled from the PCs to generate new synthetic frames. Assuming that the distribution of values along each PC captures the underlying noise properties of the data and since the PCs are uncorrelated by construction, we drew samples independently from each one (with replacements), generating new points in the space of PCs. We then applied the inverse transform and obtained synthetic intermediate frames (which follow the same noise distribution as the real frames).  {At no point in this approach did we truncate the number of PCs, thus preserving information.}

Finally, we processed these frames using the standard MATISSE DRS and obtained squared visibilities. We then repeated this process 200 times to properly sample the probability distribution. This resulted in $N=200$ noise realizations for each time-averaged visibility measurement, which we then used to calculate the correlations between visibility measurements. {In Appendix~\ref{sec:preservation_of_correlations} we show how this approach indeed preserves correlations in the observables.}

\subsection{Calculating the variance-covariance matrices}
One approach to calculate the covariance matrix is
\begin{equation}\label{eq:covmatest}
    \Sigma_{kl} = \mathbb{E}\left[(V_k - \mathbb{E}[V_k])(V_l - \mathbb{E}[V_l]) \right],
\end{equation}
where $V_k$ and $V_l$ represent two visibility measurements from a dataset ($\mathbb{E}$ denotes the expected value), and each of the indexes $k$ and $l$ correspond to unique baseline and wavelength pairs. It turns out that this estimator leads to a covariance matrix, which is invertible but not well conditioned in our case. This means that inverting this matrix significantly amplifies the estimation error. 

Therefore, as a first step, we assessed the covariance matrix $\Sigma$ using the estimator proposed in \cite{LedoitWolf}, designed to produce well-conditioned covariance matrices. We then calculated the correlation matrix, following an approach similar to that of \citet{GRAVITY_correlations}:
\begin{equation}
    K_{kl} = \frac{\Sigma_{kl}}{\sqrt{\Sigma_{kk}\Sigma_{ll}}} \, .
\end{equation}
We {then computed the} covariance matrix as  $\Sigma_{kl} = \sigma_k \sigma_l K_{kl}$, where $\sigma_k$ is the standard deviation of $V_k$ and $\sigma_l$ is the standard deviation of $V_l$, as estimated by the MATISSE data reduction pipeline. This assumes that the errors from the pipeline are correctly estimated (testing whether this is true is beyond the scope of this paper).

{In optical interferometry, calibrator stars are used to remove atmospheric and instrumental contributions (i.e., the transfer function) to observations. An ideal calibrator is a point source, as it enables the transfer function to be measured directly. However, with current optical interferometric instruments, the angular diameters of most calibrator stars are at least partially resolved. As a result, accurate knowledge of the calibrator angular diameter ($\theta_{cal} \pm \sigma_{\theta_{cal}}$) is required to compute the transfer function from calibrator observations. An uncertainty in the calibrator angular diameter ($\sigma_{\theta_{cal}}$) propagates into an uncertainty on the transfer function used to calibrate the science data. To account for this effect on the covariance matrices, we calibrated each iteration of the visibility calculated in Sect. \ref{sec:correlations_general_approach} using a different value for the calibrator angular diameter, randomly sampled from a normal distribution with mean $\theta_{cal}$ and standard deviation $\sigma_{\theta_{cal}}$.}

When calculating the covariance matrix, we assumed that the object itself did not vary during the observation ($\sim$30 min in total) and that the value of a frame at some time does not correlated with the value of a frame at another time (there is no time correlation between frames). However, we accounted for all other correlations in the data, including those between telescopes, baselines, spectral channels, and BCD positions. 

\subsection{Results}

\begin{figure}[t!]
    \centering
    \includegraphics[bb = 0 0 600 600, clip, width=0.5\textwidth]{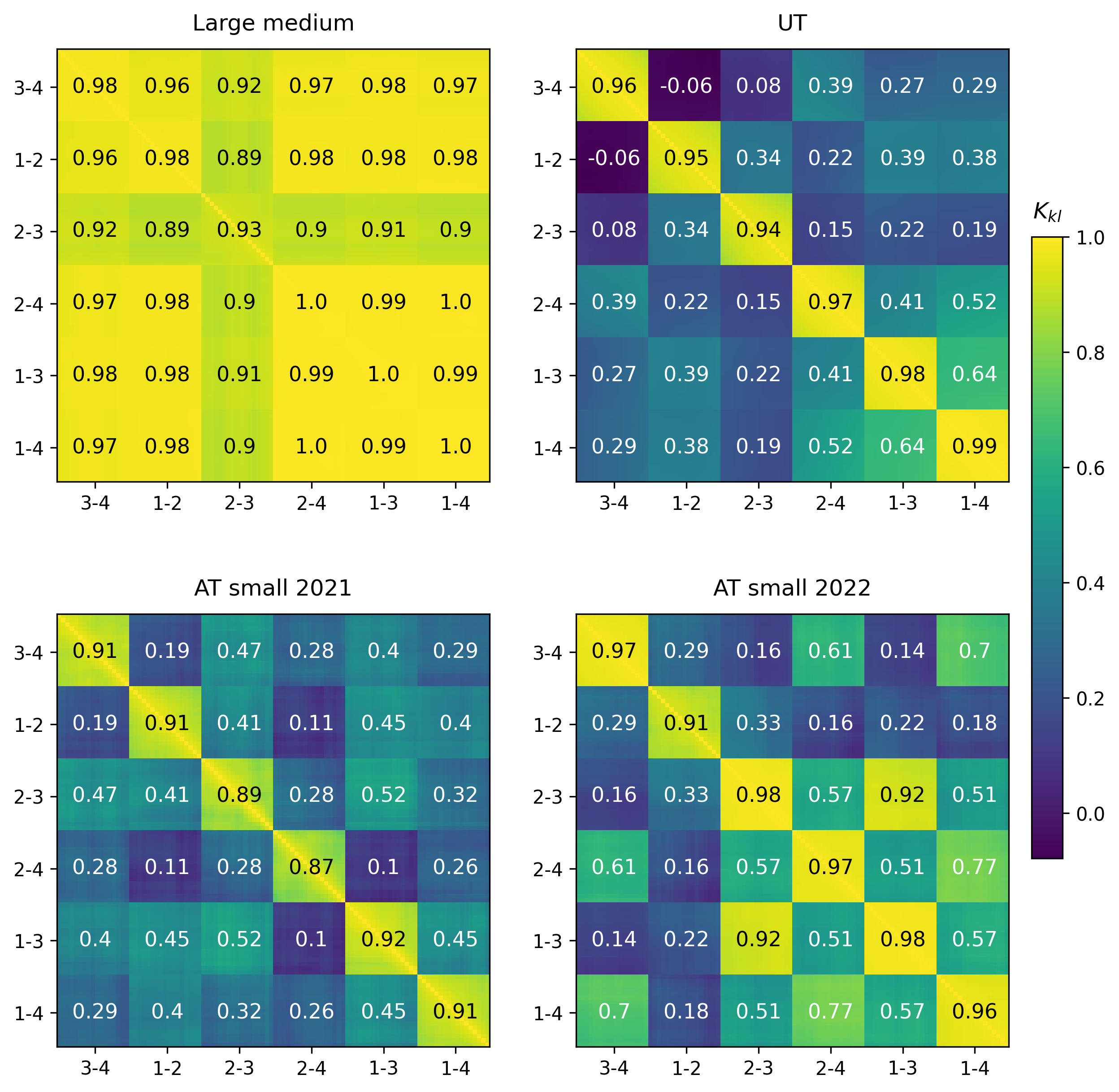}
    \caption{Correlation matrices of the $V^2$ measurements for the four MATISSE \textit{L}-band datasets (see Table~\ref{tab:observations}). 
    Each pixel in the images represents a data point for the baseline and wavelength pairs. The values on the image represent the average correlation across all spectral channels for a given baseline pair.
    \label{fig:correlationmatrixesofv2}}
\end{figure}

Figure \ref{fig:correlationmatrixesofv2} shows the correlation matrices for our four MATISSE \textit{L}-band datasets (Table ~\ref{tab:observations}). 
The six square structures along each axis correspond to different baselines, with each square structure containing 17 spectral channels between $3.2~\mu$m and $3.6~\mu$m. For most datasets the dominant correlations ($K_{kl}\sim 1$) are between spectral channels of the same baseline. This effect was also observed for GRAVITY in \cite{GRAVITY_correlations}, where the shared optical path between spectral channels is cited as the cause for these correlations. We believe the same applies for MATISSE. Furthermore, any variations in the sky thermal background emission should affect all spectral channels equally within the \textit{L} band.

For the {Large} array observations, the dominant error results from the angular size of the calibrator (HD~33042). Indeed, this calibrator is strongly resolved for this array ($0.45<V^2_{UD}<0.9$), which increases the impact of the angular size error on the total error. This results in points, which strongly correlated with one another for this dataset.

This approach shows that VLTI/MATISSE data can be highly correlated. Neglecting these correlations and assuming independent errors may therefore strongly bias the results of model fitting and parameter estimation.

\section{Model fitting} \label{sec:Models}

To further constrain the brightness distribution of the circumstellar emission, model fitting was performed on the datasets using geometric models. We considered four categories of models, with increasing degrees of complexity: (1) an over-resolved homogeneous emission (one free parameter; Sect.~\ref{sec:extended_dust}); (2) single-component emission models (four or five free parameters; Sect.~\ref{sec:spatial_info}); (3) two-component emission models (seven or eight free parameters; Sect.~\ref{sec:twocomponentmdls}); and (4) models considering a possible spectral dependence of the ratio between circumstellar emission and total flux across the \textit{L} band (one additional free parameter, included in both Sects.~\ref{sec:extended_dust} and \ref{sec:spatial_info}). For the over-resolved and single-component emission models, we systematically compared the best-fit parameters obtained with the covariance matrices calculated in Sect.~\ref{sec:correlations} to those derived under the common assumption of independent errors for each data point.

\subsection{Methodology}
\label{sec:models_method}

\subsubsection{Fitting approach}
\label{sec:fitting_approach}

To identify the model that best matches the MATISSE data, we adopted a Bayesian inference approach. For information on the log-likelihood expressions used, refer to Appendix \ref{sec:appdx_logl}.

We explored the parameter space using \texttt{nestle}\footnote{\url{https://github.com/kbarbary/nestle}}, a python implementation of the nested sampling algorithm. {Nested sampling directly estimates the Bayesian} evidence $z$, defined as $z =  \int_{}^{}\mathcal{L}(\boldsymbol{\theta})\pi(\boldsymbol{\theta})d\boldsymbol{\theta}$ where $\mathcal{L}(\boldsymbol{\theta})$ is the likelihood and $\pi(\boldsymbol{\theta})$ is the prior ({the specific priors used are described in Appendix~\ref{sec:appdx_prior}}). The higher the evidence value, the more likely the model is. {We used this quantity for model selection. To estimate the errors in the best-fit parameters, we also used a bootstrapping approach. This approach is described in Appendix~\ref{sec:appdx_logl}.}

\subsubsection{Common assumptions}
\label{sec:common_assumptions}

For all models, we represent the stellar contribution to the total $V^2$ using the limb-darkened photosphere model by \citet{limbdarkenedphotosphere} with an \textit{L}-band limb darkened coefficient $u_{\lambda, L} = 0.112$ obtained using the methodology in \citet{Claret1995} and a limb-darkened angular diameter $\theta_{\rm LD} = 0.736 \pm 0.019$ mas  \citep{Defrere_2012_betapic}. The mathematical expression for this stellar model can be found in Appendix~\ref{sec:expressions_for_V2}.
{All circumstellar models are assumed to be centered on the star (justified by the close to 0 CP).}
Our preliminary results with the geometric models described later in Sects~\ref{sec:spatial_info} and \ref{sec:twocomponentmdls} show that, in some cases, the ratio of the circumstellar flux to the total flux strongly correlates with the angular extent of the circumstellar emission. In particular, we find nonphysical solutions where the majority of the total flux in the \textit{L} band is emitted from a very compact ($\sim$~1~mas) circumstellar region. This contradicts literature measurements indicating that the flux in this band primarily originates from the star \citep[flux ratio smaller than $\sim$10\% for $\lambda \lesssim 7~\mu$m according to][]{Worthen2024}. To mitigate this effect, we used photometric measurements from WISE~W1 (centered at $\lambda = 3.4~\mu$m) coupled with a stellar model to derive an upper limit on the flux ratio. We adopted a \citet{Nextgen} NextGen stellar template ($T_{\mathrm{eff}}=8000$~K, $\log(g)=4.0$, normalized to the V-band magnitude), and used \texttt{synphot} \citep{synphot} to compute the  synthetic stellar flux in the W1 filter: $F_{\star} = 13.43$~Jy. The W1 measurement was modeled as the sum of stellar ($F_{\star}$) and circumstellar ($F_{\textrm{CSE}}$) emissions: $F_{\textrm{tot}} = F_{\star} + F_{\textrm{CSE}} = F_{\star} / (1 - f)$, where $f = F_{\textrm{CSE}} / F_{\textrm{tot}}$ is the flux ratio fitted in the models described below and evaluated at $\lambda = 3.4~\mu$m. Consequently, our fitting of MATISSE visibilities includes a comparison of $F_{\textrm{tot}}$ to the WISE W1 measurement (allWISE catalog, $12.9\pm2.1$~Jy), effectively constraining $f$ to values below 9.8\% (at 1$\sigma$).  This photometric constraint was included in the model fitting as an additional data point, without specific weighting relative to the visibility measurements.

\subsection{Homogeneous and over-resolved emission model}\label{sec:extended_dust}
\subsubsection{Models}

A common approach to model the circumstellar emission originating from hot exozodi consists in assuming that it is homogeneous and over-resolved for the shortest baselines (i.e., $V_{\mathrm{CSE}} \approx 0$, $\forall B$, where $V_{\mathrm{CSE}}$ is the visibility of the circumstellar emission). These assumptions have been employed in numerous studies of exozodi \citep[e.g.,][]{Ertel2014, Ertel2016, Absil2021} because they significantly simplify the analysis and can yield reasonable results, as demonstrated by \citet{Fomalhaut_VINCI_Absil_2009} and \citet{Defrere2011}. Under these assumptions, the observed squared visibility, $V^2(B)$, can be expressed as
\begin{equation}\label{vis_unresolved_dust}
    V^2(B) \simeq (1-f)^2 V_{\star}^2(B),
\end{equation}
where $V_{\star}$ is the stellar visibility given by Eq.~\ref{eq:Vstellar} and $f$ is the ratio between the flux of the circumstellar emission and the total flux (photospheric plus circumstellar).

We propose two variants of this model: one assuming a constant flux ratio, $f$, within the \textit{L} band (over-resolved distribution: \textbf{ORD} model) and another in which $f$ varies with wavelength according to a power law, $f\propto \lambda^{p_{\mathrm{spec}}}$, as defined in Eq.~\ref{eq:spectralslope} (\textbf{ORDS} model). The first model has a single free parameter ($f$), while the second model has two -- the flux ratio at a reference wavelength ($f_{3.4\,\mu\mathrm{m}}$) and the power-law index ($p_{\mathrm{spec}}$). Since the three 2021 datasets were obtained within a short time span ($\sim$20 days), we combined them under the assumption that the brightness distribution remains constant between observations. The 2022 dataset was excluded due to two baselines exhibiting $V^2 > 1$ and to avoid potential biases from source variability between January 2021 and October 2022. However, fitting the models both with and without the 2022 data yields consistent parameter values.

\subsubsection{Results}
The results for the best-fit flux ratio of this model to the $V^2$ measurements are presented in Table \ref{ORDresults} and shown in Fig.~\ref{fig:single_comp_fits}. 
Our combined datasets show strong detections of a circumstellar emission around $\beta$~Pictoris\ in the \textit{L} band with MATISSE. The best-fit flux ratio is around $1.8\% \pm 0.1\% $ when ignoring correlations in the data and around $2.3$--$2.9\%$ ($>7\sigma$ detection) when accounting for them, illustrating the impact of neglecting correlations on the retrieved parameters. The flux ratio appears to be higher in the \textit{L} band than in the \textit{H} band, as previously measured using VLTI/PIONIER with the same over-resolved emission model ($1.37 \pm 0.10 \pm 0.13\%$ in \citealp{Defrere_2012_betapic} and $\sim 1.4 \%$ in \citealp{Ertel2016}).

When correlations are ignored, {the data cannot distinguish between the ORD model, in which the flux ratio is wavelength-independent, and the ORDS model. In the latter, the spectral slope remains consistent with zero.} In contrast, when accounting for correlations, the model with a wavelength-dependent flux ratio provides a better fit to the data.  It reveals a flux ratio that decreases with increasing wavelength as $f \propto \lambda^{p_\text{spec}}$, with $p_\text{spec} \approx -3.2$ between $\lambda = 3.2~\mu$m and $3.6~\mu$m. In comparison, \citet{Defrere_2012_betapic} found a wavelength-independent flux ratio across the \textit{H} band. A more detailed discussion on the \textit{L}-band spectral dependence and its implications can be found in Sect.~\ref{sec:spectral_discussion}.

\begin{table}[]
    \centering
            \caption{Best-fit parameters for the over-resolved and homogeneous emission model described in Sect.~\ref{sec:extended_dust}.}

    \begin{tabular}{lcccc}
    
        \hline
        \hline

        Model & Parameter & \begin{tabular}{c} Value without \\ correlations \end{tabular} & \begin{tabular}{c} Value with \\ correlations \end{tabular} \\

        \hline

        \multirow{1}{*}{\textbf{ORD}}& $f$ [$\%$]& $1.78_{-0.12}^{+0.11}$ & $2.3_{-0.3}^{+0.2}$ \\
        & $\log(z)$ & $412$& $965$\\
        & $\chi^2_r$ / \textrm{AIC} &  $0.31$/$97$ &  $2.15$/$661$\\
        \hline        
        \multirow{2}{*}{\textbf{ORDS}}& $f_{3.4\, \mu \mathrm{m}}$ [$\%$]& $1.78_{-0.12}^{+0.11}$ & $2.9^{+0.3}_{-0.3}$ \\
        & $p_{\mathrm{spec}}$ & $-1.24_{-2.50}^{+2.49}$& $-3.17^{+0.4}_{-0.4}$\\
        & $\log(z)$ & $407.6$& $974$\\
        & $\chi^2_r$ / \textrm{AIC} & $0.31$/$97$ & $1.72$/$530$\\
        \hline
    \end{tabular}
\label{ORDresults}
\end{table}

\subsection{Single-component emission models}\label{sec:spatial_info}
One limitation of the simple model described in Sect.~\ref{sec:extended_dust} is that, if the dust is not over-resolved, the flux ratio might be misestimated, as shown in \citet{Absil2021}. Moreover, interferometric data can offer valuable spatial insights into the geometry of the circumstellar emission, which the over-resolved emission model fails to capture. To address this, we used geometric models to better constrain the brightness distribution of the circumstellar emission, beginning with single-component models.

\subsubsection{Models}
\label{subsec:2dgaussian}

Four different models were compared to the data, with their mathematical expressions for the interferometric visibility provided in Appendix \ref{sec:expressions_for_V2}. We modeled the circumstellar emission under the assumption that the dust lies in an inclined plane with a negligible vertical extent, as follows:
\begin{itemize}
    \item \textbf{2D Gaussian (G)} model: This model consists of an inclined 2D Gaussian radial profile (Eq.~\ref{eq:V2Dgaussian}).
    The free parameters include the full width at half maximum (FWHM) of the Gaussian {along the major axis}, the inclination $i_\text{tilt}$,
    the position angle (PA) of the apparent major axis, and the ratio between the circumstellar and the total emissions ($f$). 
    \item \textbf{2D Gaussian spectral (GS)} model: This model is similar to the 2D Gaussian model above but with $f$ varying with wavelength following a power law with an index $p_\text{spec}$ (Eq.~\ref{eq:spectralslope}).
    \item \textbf{Gaussian ring (GR)} model: This model consists of an inclined ring with a Gaussian radial profile (Eq.~\ref{eq:Vgaussianring}). The free parameters include the radius of the ring ($r_\text{ring}$), the FWHM of the Gaussian profile of the ring ($\textrm{w}_{\textrm{FWHM}}$), the inclination ($i_\text{tilt}$), the PA of the apparent major axis and the circumstellar to total flux ratio ($f$).
    \item \textbf{Gaussian ring spectral (GRS)} model: This model is similar to the Gaussian ring model above but with $f$ varying with wavelength following a power law with index $p_\text{spec}$ (Eq.~\ref{eq:spectralslope}).
\end{itemize}

\subsubsection{Results}

\begin{table}[]
    \centering
            \caption{Best-fit parameters for four single-component emission models described in Sect.~\ref{sec:spatial_info}.} \label{tab:models_one_component}
    \begin{tabular}{lccc}
    
        \hline
        \hline

        Model & Parameter & \begin{tabular}{c} Value without \\ correlations \end{tabular} & \begin{tabular}{c} Value with \\ correlations \end{tabular} \\

        \hline
        
        \multirow{4}{4em}{\textbf{2D Gaussian (G)}}&  FWHM [au]& $7.0^{+0.4}_{-3.1}$ & $0.08_{-0.004}^{+0.005}$ \\
        &$i_{\textrm{tilt}}$ [deg]& $89.7^{+0.2}_{-0.4}$ & $89.4_{-0.1}^{+0.1}$\\
        &PA [deg]& $-67.8^{+33.8}_{-0.3}$ & $-29.3_{+0.9}^{-1.0}$\\
        &$f$ [$\%$]& $2.3^{+0.1}_{-0.2}$& $6.1_{+0.3}^{-0.3}$\\
        &$\log(z)$ & $400.5$& $1032$\\
        &$\chi^2_r$ / \textrm{AIC} & $0.24$/$82$ & $0.60$/$189$\\
        
        \hline
        \multirow{4}{4em}{\textbf{2D Gaussian spectral (GS)}}& FWHM [au]& $4.5^{+1.4}_{-1.1}$ & $0.089_{-0.006}^{+0.005}$\\
        &$i_{\textrm{tilt}}$ [deg]& $89.61^{+0.94}_{-0.3}$& $68_{-3}^{+4}$\\
        &PA [deg]& $-68.0^{+34.0}_{-0.01}$& $-27_{-1}^{+1}$\\
        &$f_{3.4\, \mu \mathrm{m}}$ [$\%$]& $2.3^{+0.1}_{-0.1}$ & $7.3_{-0.6}^{+0.7}$\\
        &$p_\text{spec}$ & $-1.5^{+1.6}_{-1.8}$& $-2.2_{-0.2}^{+0.2}$\\
        &$\log(z)$ & $405.0$& $1045$\\
        &$\chi^2_r$ / \textrm{AIC} & $0.24$/$83$ & $0.41$/$134$\\
        
        \hline        
        \multirow{4}{4em}{\textbf{Gaussian ring (GR)}} & $r_\text{ring}$ [au]& $1.9^{+1.2}_{-0.6}$& $0.057_{-0.004}^{+0.004}$ \\
        & $\textrm{w}_{\textrm{FWHM}}$ [au]& $0.012^{+0.012}_{-0.004}$& $0.003_{-0.001}^{+0.001}$\\
        & $i_{\textrm{tilt}}$ [deg]& $89.8^{+0.1}_{-1.0}$& $88.8_{-0.4}^{+0.2}$\\
        & PA [deg]& $-67^{+34}_{-2}$& $-29.5_{-0.9}^{+0.8}$\\
        & $f$ [$\%$]& $2.3^{+0.1}_{-0.1}$& $3.9_{-0.3}^{+0.3}$\\
        &$\log(z)$ & $400.5$& $1029$\\
        &$\chi^2_r$ / \textrm{AIC} & $0.22$/$76$ & $0.59$/$190$\\

        \hline
        \multirow{4}{4em}{\textbf{Gaussian ring spectral (GRS)}} & $r_\text{ring}$ [au]& $1.7^{+1.1}_{-0.6}$& $0.062_{-0.005}^{+0.005}$ \\
        & $\textrm{w}_{\textrm{FWHM}}$ [au]& $0.015^{+0.013}_{-0.007}$ & $0.0027_{-0.0004}^{+0.0004}$\\
        & $i_{\textrm{tilt}}$ [deg]& $89.8^{+0.2}_{-1.2}$& $63.0_{-4.0}^{+5.0}$\\
        & PA [deg]& $-66^{+33}_{-4}$& $-27_{-1}^{+1}$\\
        & $f$ [$\%$]& $2.3^{+0.2}_{-0.1}$& $5.3_{-0.6}^{+0.5}$\\
        &$p_\text{spec}$ & $-1.3^{+1.5}_{-1.7}$& $-2.1_{-0.2}^{+0.3}$\\

        & $\log(z)$ & $394$ & $1044$\\
         &  $\chi^2_r$ /  \textrm{AIC} &  $0.22$/$78$ &  $0.41$/$136$\\
        \hline
    \end{tabular}
    \tablefoot{
The $\log(z)$ values were computed using the parameters that provide the best fit to the data. The AIC accounts for the quality of the fit and penalizes for the complexity of the model (see Eq.~\ref{eq:AIC}).}
\end{table}

\begin{table}[]
    \centering
            \caption{Best-fit parameters for the two-component emission models described in Sect.~\ref{sec:twocomponentmdls}.}

    \begin{tabular}{lcccc}
    
        \hline
        \hline

        Model & Parameter & \begin{tabular}{c} Value with \\ correlations \end{tabular}  \\

        \hline
        
        \multirow{3}{7em}{\textbf{Inner 2D Gaussian  \\ + outer ring (GIRO) }} & FWHM$_\text{inner}$ [au]& $0.067^{+0.004}_{-0.003}$ \\
        &  $i_{\textrm{tilt}}$ [deg]& $88^{+2}_{-8}$\\
        &  $PA$ [deg]& $-28.1^{+2.4}_{-0.8}$\\
        &  $f$ [$\%$] & $6.4^{+0.6}_{-0.6}$\\
        \cline{2-3}
        &  $r_\text{outer}$ [au]& $0.9^{+0.3}_{-0.3}$\\
        & $\textrm{w}_{\textrm{FWHM, outer}}$ [au]& $0.007^{+0.004}_{-0.002}$\\
        &  $f_\text{outer}$ [$\%$]& $1.5^{+0.1}_{-0.1}$\\

        & $\log(z)$ & $1029$\\
        & $\chi^2_r$ / \textrm{AIC} & $0.45$/$150$\\

        \hline      
        \multirow{3}{7em}{\textbf{Two Gaussian \\ rings (RIRO)}} & $r_\text{ring}$ [au]& $0.045^{+0.003}_{-0.003}$ \\
        &  $i_{\textrm{tilt}}$ [deg]& $81^{+5}_{-9}$\\
        &  $PA$ [deg]& $-27^{+2}_{-1}$\\
        &  $f$ [$\%$] & $4.4^{+0.5}_{-0.7}$\\
        \cline{2-3}
        & $r_\text{outer}$ [au]& $0.9^{+0.3}_{-0.2}$\\
        &  $\textrm{w}_{\textrm{FWHM, outer}}$ [au]& $0.006^{+0.002}_{-0.003}$\\
        &  $f_\text{outer}$ [$\%$]& $1.6^{+0.2}_{-0.3}$\\

        & $\log(z)$& $1037$\\
        & $\chi^2_r$ / \textrm{AIC} & $0.44$/$147$\\
        \hline  
    \end{tabular}
\label{twocompresults}
\end{table}

The model results are summarized in Table~\ref{tab:models_one_component} and presented in Fig.~\ref{fig:single_comp_fits}, with the posterior distributions shown in Figs.~\ref{fig:posterior_dists_incomp_onecomp} and \ref{fig:posterior_spectral_index}. When correlations are ignored, we find that the single-component models do not fit the $V^2$ measurements better than the homogeneous over-resolved emission model described in Sect.~\ref{sec:extended_dust} (similar $\log(z)$ values). They consistently result in circumstellar emission with a flux ratio ($f$ or $f_{3.4\, \mu \mathrm{m}}$) of about 2.3\%, extending over a few au, viewed nearly edge-on ($i_\text{tilt} \sim 89\degree$) at a loosely constrained PA between approximately $-30\degree$ and $-70\degree$.

However, when accounting for correlations in the data, the geometric models provide significant improvement in the fit compared to the over-resolved model, {although the data cannot reliably distinguish between the Gaussian ring and the 2D-Gaussian models.} For all models with correlations, we find a compact emission at approximately 0.05~au ($\sim$7 stellar radii or 2.6 mas) from the star, oriented edge-on ($i_\text{tilt} \sim 89^\degree$) when considering the case of a constant flux ratio, with a significantly better-constrained PA misaligned with the outer disk (PA $\approx -30\degree$ with an uncertainty of a few degrees). In this case, accounting for correlations gives more weight to the datasets at higher spatial frequencies (AT Large Medium and UT data) compared to those at smaller spatial frequencies (AT small data), which explains why the best-fit model is more compact. Furthermore, due to the strongly correlated noise between baselines at different orientations, the PA is more strongly constrained. Accounting for the spectral dependence of the flux ratio yields a lower inclination ($i_\text{tilt} \approx 65\degree$), which may reflect a limitation of the flat geometric models, where vertical structure in the emission is reproduced by artificially lowering the inclination.

It is noteworthy that, when accounting for correlations in the data, the inferred flux ratio depends on the assumed geometry of the circumstellar emission, ranging from approximately 6\% (G model) to 7.5\% (GS) for the 2D Gaussian model, and from about 3.9\% (GR) to 5.3\% (GRS) for the Gaussian ring model. 
{Furthermore, when ignoring correlations, introducing a spectral dependence on the flux ratio is not justified {in light of the $\log(z)$ and the Akaike information criterion (AIC).}} In contrast, when correlations are included, the spectral dependence leads to a significantly improved fit, with a best-fit index $p_\text{spec} \approx -2.2$ and an uncertainty of about $10\%$.  However, as with the constant-flux-ratio models, the 2D Gaussian and Gaussian ring geometries remain statistically indistinguishable.

It is important to note that, when correlations in the data are accounted for, the radial extent of the emission is nearly at the angular resolution limit of the Large-Medium and UT arrays. As a result, distinguishing between different geometries -- and even more so, resolving the inner edge of the emission, if it exists -- may be challenging. Additionally, when accounting for correlations, the model systematically overestimates the visibilities at low spatial frequencies (i.e., those from the AT Small array) compared to the observations (see middle and bottom panels of Fig.~\ref{fig:single_comp_fits}). This {could}suggest a more extended emission being resolved within the interferometer's FOV. Consequently, we {explored} a two-component model described in the next subsection.

\subsection{A two-component emission model} \label{sec:twocomponentmdls}

\subsubsection{Models}

As discussed in Sect.~\ref{sec:spatial_info}, when accounting for correlations in the data, the models successfully reproduce the visibilities at high spatial frequencies (and their slope with $B/\lambda$) but tend overestimate the visibility at lower spatial frequencies. Given that a compact emission seems to fit the data better according to the Bayesian evidence ($\log(z)$), we propose in this section a two-component emission model in an effort to better reproduce the lower spatial frequencies. We compare the $V^2$ measurements to two models following the same approach as that in Sect.~\ref{sec:spatial_info}, restricting our analysis to the case where correlations are accounted for in order to maintain a high $\log(z)$ value.

We used the 2D Gaussian and Gaussian ring models from Sect.~\ref{sec:spatial_info} as the inner component, combining each with an outer Gaussian ring. The free parameters for the outer component consist of the ring position ($r_\text{outer}$), the FWHM of the convolving Gaussian ($\textrm{w}_{\textrm{FWHM, outer}}$), and the flux ratio ($f_\text{outer}$). To minimize degeneracies when using the Gaussian ring model for the inner component, we fixed its width to zero (infinitely thin ring), according to the best fit for the single-component models. We also assumed that the inner and outer components share the same inclination ($i_\text{tilt}$) and PA to minimize degeneracies.

\subsubsection{Results}

The model results are summarized in Table~\ref{twocompresults} and presented in Fig.~\ref{fig:two_comp_fits}, with the posterior distributions shown in Figs.~\ref{fig:posterior_dists_incomp_onecomp} and \ref{fig:post_Rhl_outer}. {We note that the best fit to the data is for a single-component model with a spectral dependency on the flux ratio. However, the two-component models yield fits of similar quality to the nonspectral versions of the single-component models (in light of the log-evidence).}
Most parameters for the inner disk (such as the distance to the star, flux ratio, and PA) are very close to those of the one-component emission model, except for the inclination $i_\text{tilt}$, which is a few degrees lower. This could be due to the emission becoming vertically thicker with increasing distance from the star, a feature that might be better captured by reducing the inclination of the flat circumstellar emission model.
 
As expected, the addition of an external ring improves the fit to the lowest spatial frequency data. The best-fit models suggest that this outer ring is radially thin and located near 1~au (51~mas). It contributes approximately 1.5\% of the total flux in the \textit{L} band, which is one third as bright as the emission from the inner component in the two-Gaussian ring model. We note that fixing the PA of the outer component to $30 \degree$ (i.e., that of the outer known cold disk) does not significantly change the quality of the fit. In our models, the fitted PA is dominated by the PA signature from the inner component.

\section{Implications for the hot dust properties}\label{sec:discussion}

The MATISSE \textit{L}-band observations of $\beta$~Pictoris' circumstellar emission are best explained by a compact emitting region at $\sim 0.05$~au, highly inclined (nearly edge-on in most models) and misaligned by $\sim 60\degree$ on the sky plane relative to the outer disk (see Fig.~\ref{fig:posterior_dists_incomp_onecomp} for a summary of the main findings). Depending on the geometric model chosen, the flux ratio ranges from $\sim 4\%$ to $\sim 7.5\%$ at $\lambda = 3.4~\mu$m. Additionally, an outer component located close to $\sim 1~$au {could} be present in the data. We interpret the \textit{L}-band emission as originating from circumstellar dust.

{These results are broadly consistent with previous MATISSE studies of hot-dust systems, such as $\kappa$ Tuc \citep{Kirchschlager2020} and Fomalhaut \citep{Ollmann2025}, in which the emission appears compact, compatible with dust located near the sublimation region, and likely dominated by submicron-sized grains. However, our analysis differs in two important respects. First, we developed a method to explicitly estimate and incorporate the full covariance matrix of the MATISSE data into the fitting approach, showing that neglecting correlations can significantly bias the inferred parameters. Second, this study represents the first case in which MATISSE data alone directly constrain the spatial extent and orientation of the hot-dust emission.}

\subsection{On the spatial information} \label{sec:onthespatialinfo}

Models of the spectral energy distribution (SED) of hot exozodis, primarily constrained by the near- to mid-infrared flux ratio, have consistently indicated that the hot dust is concentrated in the innermost regions of planetary systems, near the sublimation radius \citep[e.g.,][]{Absil2006, Absil2008, Difolco2007, Defrere2011, Lebreton2013, Kirchschlager2017}. Our MATISSE observations enable a direct test of this hypothesis for $\beta$~Pictoris.

\begin{figure}[t!]
    \centering

    \includegraphics[bb = 5 5 430 396, clip, width=0.43\textwidth]{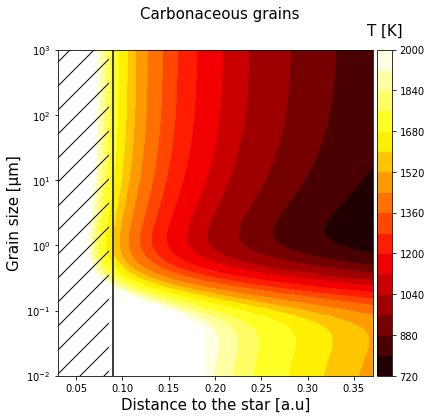}
    \caption{Temperature of carbonaceous dust grains as a function of their size and distance to the star, assuming hard spheres. The white region marks the sublimation zone ($T>2000$~K), while the shaded area indicates the possible location of the half-flux radius of the emission detected with MATISSE.}
    \label{fig:temperatures}
\end{figure}

To investigate this further, we computed the equilibrium temperature of carbonaceous grains \citep{zubko1996}, a composition favored in these previous studies, using Mie theory under the assumption of spherical, homogeneous particles and adopting the stellar spectrum described in Sect.~\ref{sec:common_assumptions}. Figure~\ref{fig:temperatures} shows the resulting grain temperatures as a function of grain size and distance from the star, alongside the spatial constraints derived from the MATISSE observations (Sect.~\ref{sec:spatial_info}). These results indicate that the emission arises from a region located at or within the expected sublimation radius, corresponding to dust temperatures  approaching or exceeding 2000~K.

Therefore, our MATISSE data of $\beta$~Pictoris\ provide, for the first time, direct observational evidence{, for any system,} supporting a scenario in which the hot dust resides near the sublimation region in agreement with predictions from SED-based models. They also disfavor an alternative scenario in which very small grains, located at larger stellocentric distances and out of thermal equilibrium (i.e., stochastically heated by individual photons), mimic the near- to mid-infrared flux ratios expected from thermally equilibrated grains situated closer to the star. 

{Previous studies typically constrained the dust to be within the FOV of the instrument. This, however, does not mean that it is located at the sublimation radius, as the FOV is much larger than this radius for all the systems studied previously (typically one to a few times 0.1 arcsec).}

\subsection{On the spectral slope} \label{sec:spectral_discussion}

VLTI/PIONIER observations of $\beta$~Pictoris\ revealed a wavelength-independent flux ratio in the \textit{H} band, suggestive of scattering-dominated emission, with up to 70\% potentially arising from forward or backward scattering by dust grains in the outer, edge-on debris disk \citep{Defrere_2012_betapic}. In contrast, our \textit{L}-band MATISSE observations reveal a wavelength-dependent flux ratio, described by a power law $f \propto \lambda^{p_\text{spec}}$ over the 3.2–3.6~$\mu$m wavelength range with a spectral index $p_\text{spec} = -2.2 \pm 0.2$.

To assess the implications for the grain properties, we performed first-order modeling of the circumstellar emission, $F_{\rm CSE}$, assuming it arises from a population of grains with a single characteristic size. In this approximation, the circumstellar emission is given by $F_{\rm CSE} \propto Q_{\rm abs}(2\pi s / \lambda)\times B_{\nu}(T)$, where $Q_{\rm abs}$ is the absorption efficiency of grains with radius $s$ at wavelength $\lambda = c/\nu$, and $B_{\nu}(T)$ is the Planck function at dust temperature $T$. Further assuming power-law behaviors within the relevant wavelength range (3.2–3.6~$\mu$m) such that $Q_{\rm abs} \propto \lambda^{p_{\rm abs}}$, $B_\nu(T) \propto \lambda^{p_{\rm BB}}$, and $F_{\star} \propto \lambda^{p_{\star}}$, the flux ratio becomes
\begin{equation}
    f = \frac{F_{\rm CSE}}{F_{\star} + F_{\rm CSE}} \approx \frac{F_{\rm CSE}}{F_{\star}} \propto \frac{\lambda^{p_{\rm abs} + p_{\rm BB}}}{\lambda^{p_{\star}}} = \lambda^{p_{\rm spec}} .
\end{equation}
This yields the relation $p_{\rm abs} = p_{\rm spec} + p_{\star} - p_{\rm BB}$.
Using the stellar model described in Sect. \ref{sec:common_assumptions}, we find $p_{\star} = -1.80$, yielding $p_{\rm abs} = - 4 \pm 0.2 - p_{\rm BB}$ between $\lambda = 3.2$ and $3.6~\mu$m.

\begin{figure}[t!]
    \centering
    \includegraphics[width=0.45\textwidth]{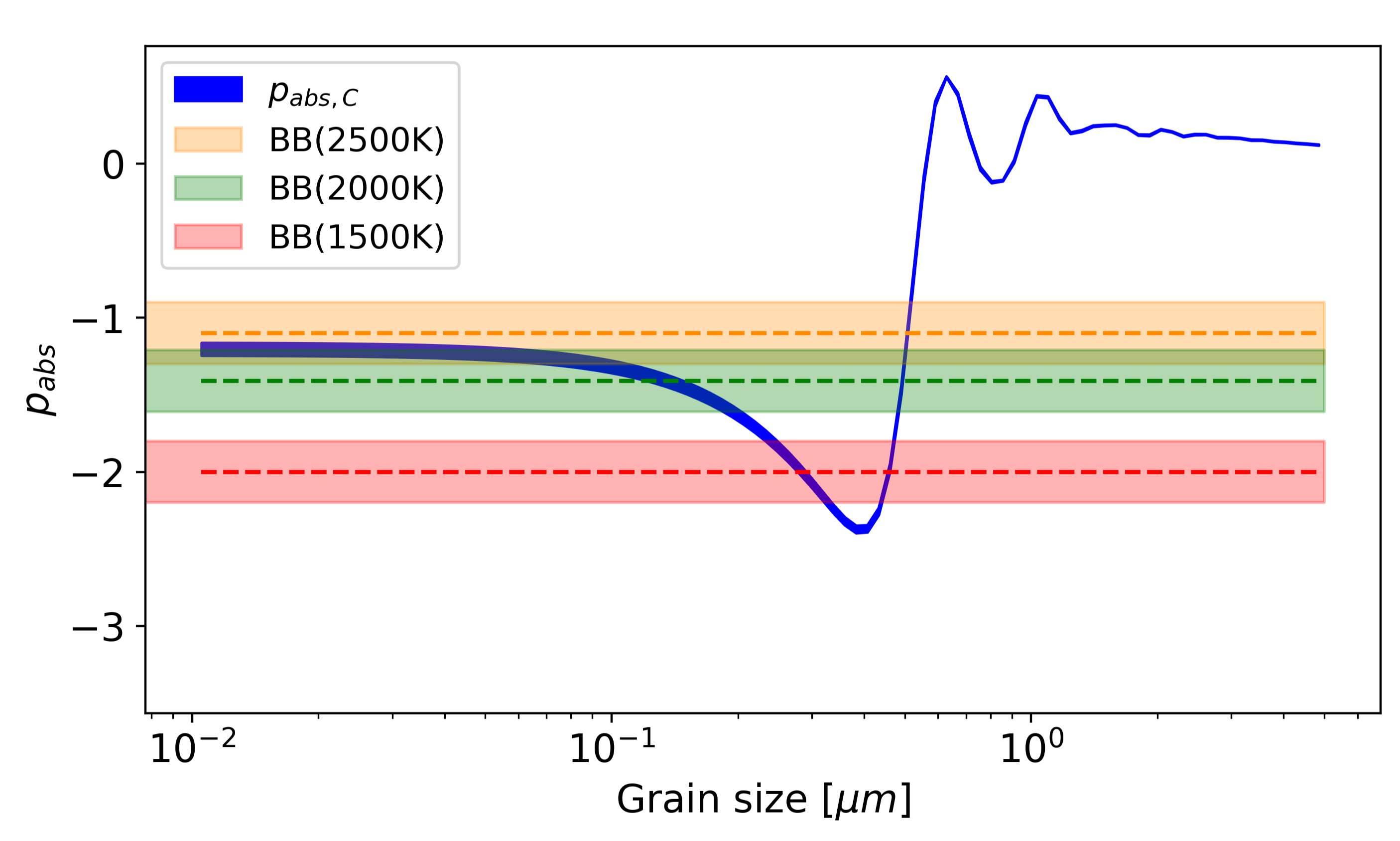}
    \caption{Power-law exponent ($p_{\rm abs}$) of the absorption efficiency, with corresponding 1$\sigma$ error, for carbonaceous grains between $\lambda = 3.2$ and $3.6~\mu$m as a function of grain size.
    Measured values (GS model; Sect.~\ref{sec:twocomponentmdls}) assuming three BB temperatures 
    are shown, with shaded areas indicating 1$\sigma$ uncertainties.
    \label{fig:slope_G}}
\end{figure}

Therefore, assuming a blackbody (BB) temperature $T$, we can derive the spectral power-law exponent of the absorption efficiency ($p_{\rm abs}$) in the \textit{L} band from the observed data and compare it to the theoretical predictions from Mie theory for various grain sizes. As shown in Fig.~\ref{fig:slope_G} for carbonaceous grains, the \textit{L}-band emission detected with MATISSE appears to be dominated by submicron-sized particles (i.e., $\lesssim 0.5~\mu$m), consistent with grains in the Rayleigh regime (where $s \lesssim \lambda / 2 \pi$), and in line with predictions from radiative transfer models fitting the SED of hot exozodis.

However, this analysis assumes emission from a single grain at a single temperature, whereas the hot dust likely spans a range of sizes and temperatures. Moreover, the relative contribution of scattered light versus thermal emission remains uncertain. As discussed in \citet{Defrere_2012_betapic}, scattering may dominate in the \textit{H} band, while thermal emission prevails in the \textit{L} band. A detailed investigation of the spectral properties of this hot exozodi, possibly affected by temporal variability, is beyond the scope of this paper and is deferred to future work.

\subsection{On the origin of the hot dust}

Our MATISSE observations of $\beta$~Pictoris\ locate the hot exozodi near the sublimation distance and indicate that it is composed of submicron-sized grains. Maintaining such levels of hot dust in planetary systems is a challenge \citep[see reviews in][]{Kral2017, ertel2025}, and our findings may lend additional support to the exocometary delivery scenario, in which inward-scattered comets release dust as they sublimate near the star.

Exocometary activity around $\beta$~Pictoris\ has been studied for over three decades, initially through transient absorption features in metallic lines \citep[e.g., \ion{Ca}{ii} and \ion{Mg}{ii};][]{Beust1990, Beust1996} and more recently through continuum photometric detections of exocomets transiting the star \citep{Zieba2019, Lecavelier2022}. The variability in mid-infrared emission (5–15~$\mu$m) from hot and warm dust over a 20-year timescale may also be linked to exocometary activity \citep{Chen2024}. Dynamical models attribute the anisotropic distribution of exocomet infall to high-order mean-motion resonances with a giant planet \citep{Beust1996, Beust2000}. This mechanism was recently revisited by \citet{Beust2024}, who identified a stable reservoir of exocomets near 1~au likely perturbed by $\beta$~Pictoris\ c. Our MATISSE results appear compatible with both a directional clustering of exocomet activity and the presence of a dust component near 1~au.

Dynamical models propose that star-grazing exocomets originate from planetesimals initially near the disk midplane, with low inclinations. Resonant interactions with a planet excite their eccentricities \citep[e.g.,][]{Faramaz2017}, sending them on star-grazing orbits and making them observable when crossing the line of sight near the dust sublimation zone. \citet{Beust2007} further showed that large inclination oscillations (up to several tens of degrees) can occur specifically when the bodies become star-grazers, driven by Kozai-Lidov–like effects within the mean-motion resonance.
This mechanism may help explain the observed $\sim$60° misalignment of the hot dust emission relative to the outer disk. As shown by \citet{Beust2007}, exocomets reach their maximum inclination when their argument of periastron $\omega$ is near 0° or 180°, meaning that their orbits remain aligned in an apsidal direction with the disk midplane, while their orbital planes tilt significantly. Although the comets themselves do not stray far from the midplane near periastron, their dust byproducts do due to radiation pressure, which alters their orbital parameters, particularly $\omega$, without changing their orbital plane orientation (inclination and node). As a result, the dust inherits the inclined geometry of the parent orbit but decouples dynamically due to radiation pressure, potentially producing the observed misaligned emission. Further modeling is required to test this hypothesis.

{However, from a theoretical standpoint, the cometary scenario by itself cannot explain the observed hot exozodis. The deposited dust should quickly be removed from this region either by sublimating, colliding, or blowing away under radiation pressure \citep{ertel2025}. This model would require unreasonably high dust deposition rates to compensate for the short grain lifetimes. Furthermore, \cite{Pearce2022} showed that explaining the ratio between near- and mid-infrared emission, particularly the lack of mid-infrared emission in hot dust systems, would require the dust to be deposited with an unphysically steep size distribution. The cometary scenario would therefore require a trapping mechanism to successfully explain the observations.}
\section{Implications for observing circumstellar dust} \label{sec:discussion_interfero}

\subsection{Apparent variability induced by discrete $uv$ sampling} \label{sec:u-vsamplingandfrat}

Previous studies of hot exozodis often assumed over-resolved, homogeneous emission, a necessary simplification given past observational limitations. However, discussing variability on the basis of this assumption involves some risks. Indeed, if the emission is not truly over-resolved, changes in $uv$ coverage can lead to apparent variations in the retrieved flux ratio under that assumption. To test this, we modeled the emission as a 2D Gaussian with known geometric parameters, ensuring that it is not over-resolved under typical observational conditions. We then retrieved the flux ratio assuming that the emission is over-resolved using the $uv$ plane at different hour angles. We estimated the errors via bootstrapping, where each sample consists of all spectral channels for one baseline. 

\begin{figure}[t!]
    \centering
    \includegraphics[width=0.4\textwidth]{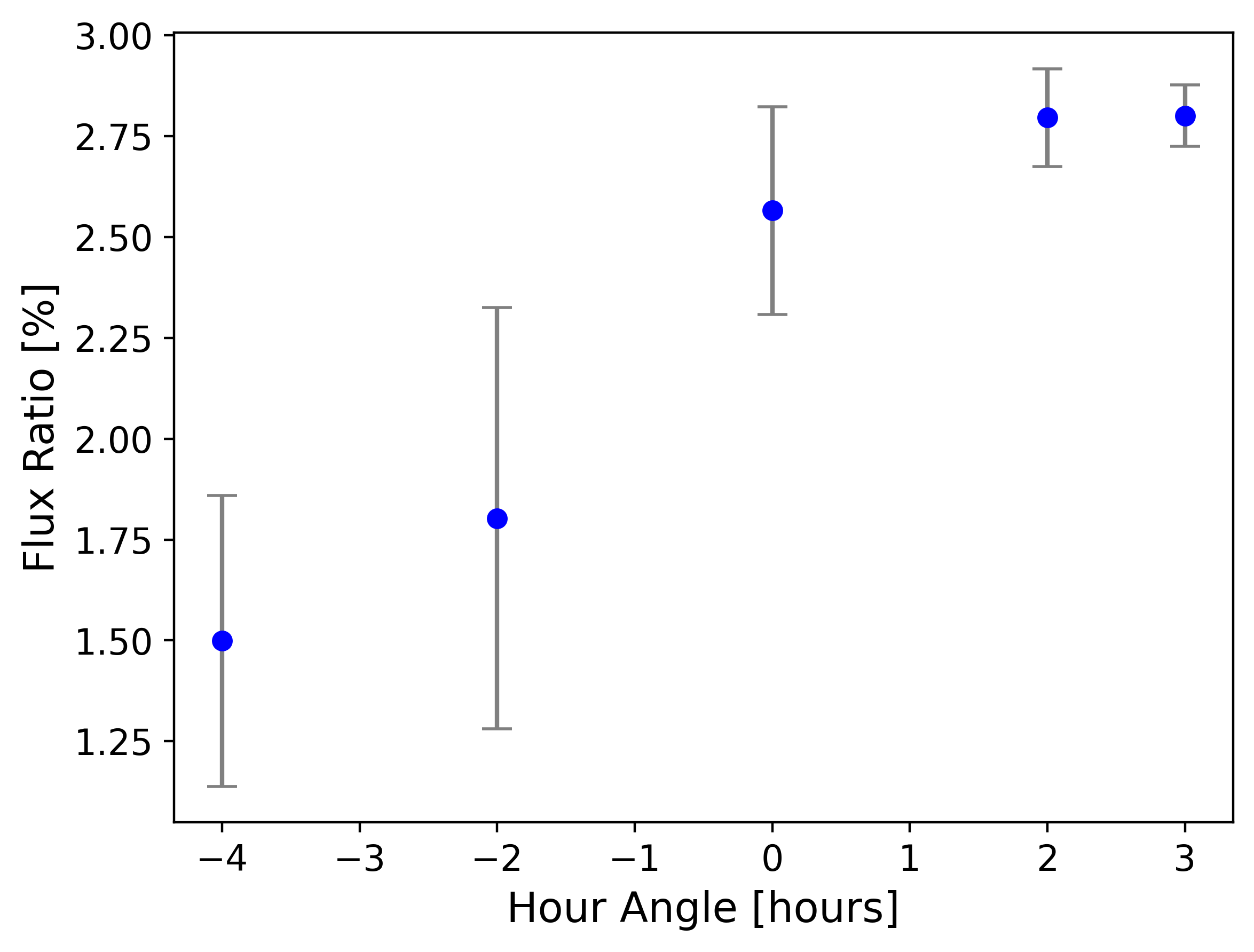}
    \caption{Example of retrieved flux ratios from a synthetic 2D Gaussian model, assuming over-resolved emission. Variations due to different $uv$-plane samplings could mimic temporal variability. \label{fig:variability}}
\end{figure}

We find that, for certain geometries of the emission and baseline configurations (Medium, UT), variations in hour angle can result in changes in the retrieved flux ratio, which may be misinterpreted as variability. To illustrate this effect, Fig.~\ref{fig:variability} shows the retrieved flux ratios as a function of hour angle for an input model with parameters: FWHM = 17~mas, $i_{\rm tilt} = 89.9\degree$, PA = $-80\degree$, and $f = 3\%$. Two observations at different hour angles could be misinterpreted as variable emission, while the effect results from incorrectly assuming an over-resolved emission combined with a different sampling of the $uv$ plane. We notice that this effect is most pronounced when the $uv$ plane is predominantly sampled in one direction. Notably, the retrieved flux ratio differs by more than $3\sigma$ between hour angles of –4 and 3.

\subsection{On the importance of taking into account correlations}\label{sec:correlations_for_bright}
{Previous studies have shown the impact of accounting for correlations in the model fitting of interferometric data. \cite{Lachaume2019} showed that ignoring correlations has a strong impact on the estimated stellar uniform disk diameters and their uncertainty observed with VLTI/PIONIER. \cite{Lachaume2021} quantified the bias introduced by ignoring correlations in interferometric model fitting. We also see this effect in the present study.} As shown in Sect.~\ref{sec:Models}, accounting for correlations has a significant impact on the modeling results. While the circumstellar emission around $\beta$~Pictoris\ is admittedly faint, we argue that neglecting correlations can also influence the analysis of much brighter objects.

To illustrate this, we generated synthetic datasets using the covariance matrices calculated in Sect.~\ref{sec:correlations}, simulating an object with a known spatial distribution (2D, PA~$=-40\degree$, $i_{\rm tilt}=30\degree$, FWHM~=~12~mas), and tested a range of flux ratios from faint ($f \approx 5\%$) to very bright ($f \approx 85\%$). We generated synthetic datasets with multiple noise realizations ($\sim$200) and compared the input parameters to those retrieved using the same methodology in Sect.~\ref{sec:fitting_approach}. Here, we (i) assumed uncorrelated (independent) errors and (ii) accounted for correlations in the model fitting. {Using these two assumptions, we furthermore compared the errors obtained with the nested sampling algorithm.} It is important to note that this approach supposes that the correlation matrices are perfectly estimated, as the same matrix was used both to generate the synthetic data and to retrieve the model parameters. 

\begin{figure*}[tbph!]
    \centering
    \includegraphics[width=\textwidth]{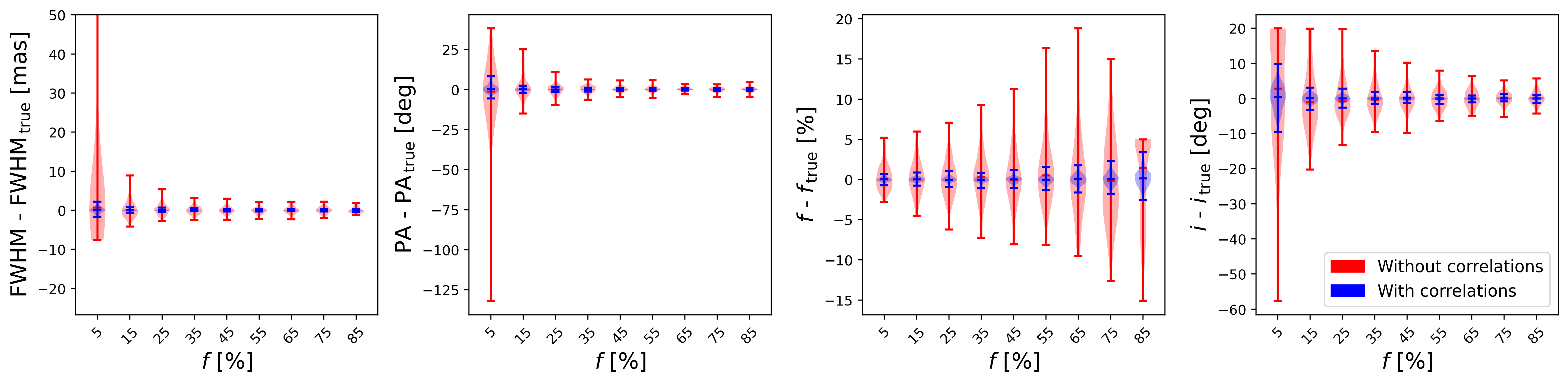}
    
    \vspace{0.5cm}
    
    \includegraphics[width=\textwidth]{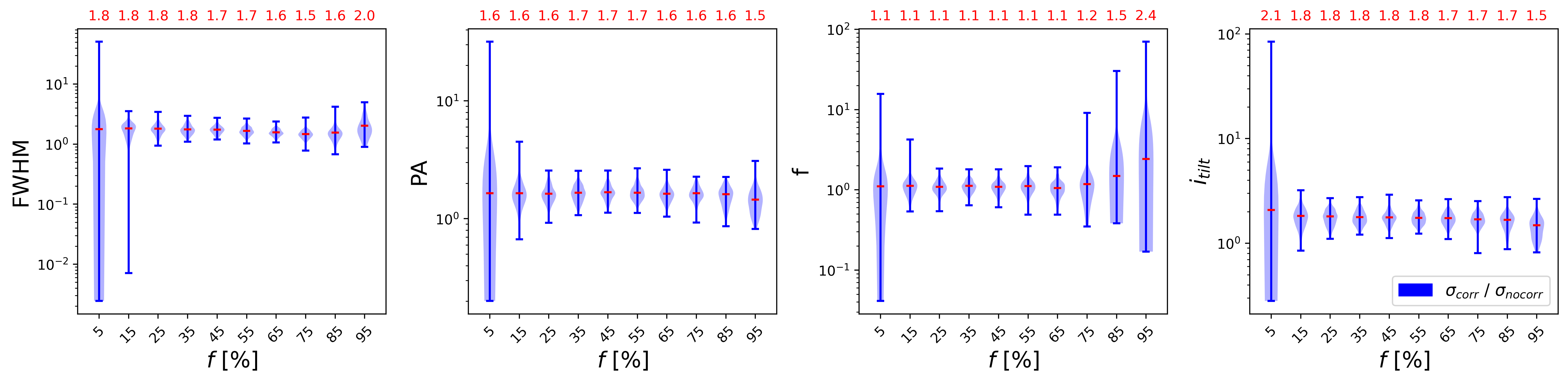}
    
    \caption{\textbf{Top:} Distribution of the difference between the retrieved values and the input value for four parameters (FWHM, PA, flux ratio, and inclination). The red distributions were obtained under the assumption of independent errors, while the blue distributions account for correlations. These distributions are shown in the form of violin plots with increasing flux ratios on the horizontal axis. In each case, the vertical markers indicate the minimum, median, and maximum values. \textbf{Bottom:} Distribution of the ratios between the errors when accounting for correlations and errors when ignoring correlations. The median value is shown by the red text at the top of each violin plot.}
    
    \label{fig:violin_bright}
\end{figure*}

The top panels of Fig.~\ref{fig:violin_bright} present the distributions of the differences between the retrieved and input values for the four model parameters (PA, $i_{\rm tilt}$, FWHM, and $f$), as a function of flux ratio ($f$). These results, obtained over multiple noise realizations, show that accounting for correlations significantly reduces the dispersion around the true values, even for bright sources. For example, with $f \approx 50\%$, the {bias} on the retrieved parameters is $\sim 6$ times smaller when correlations are accounted for. The bottom panels of Fig.~\ref{fig:violin_bright} present the distributions of the ratio between the error when accounting for correlations and the error when ignoring them. 

The median of this ratio is bigger than that for all parameters, meaning that when accounting for correlations, the errors on the parameters are bigger than when ignoring correlations. This conclusion was also found by \citep{Lachaume2019} using PIONIER for stellar diameter estimates. This highlights the importance of properly accounting for correlations in VLTI/MATISSE data to improve the precision of the retrieved parameters.

\section{Conclusions}
In this study, we presented \textit{L}-band observations of $\beta$~Pictoris\ obtained with MATISSE, a mid-infrared interferometric instrument operating in the \textit{L}, \textit{M}, and \textit{N} bands at the VLTI, aiming to probe the inner regions of the circumstellar environment. The data reveal emission consistent with hot exozodi. 

We introduced a new method to estimate correlations between visibility measurements and found strong wavelength correlations within a given baseline. Additionally, we showed that uncertainties in the calibrator's angular size can induce significant correlations across all measurements. We modeled the hot exozodi emission of $\beta$~Pictoris\ and compared our results obtained with and without accounting for correlations -- a common simplification. Our analysis shows that including these correlations significantly affects the inferred parameters, which is crucial for faint sources such as hot exozodis, and may also influence results for brighter targets. For example, when assuming independent errors, the spectral slope of the flux ratio is consistent with zero; however, when correlations are accounted for, we find that the flux ratio decreases with wavelength. This is the first time correlations have been measured and accounted for MATISSE data. This approach can be integrated into the standard MATISSE pipeline, using the OIFITS 2 format \citep{OIFITS2}, although particular care should be taken for highly correlated datasets where the covariance matrix might be singular (determinant close to 0 and therefore unstable under inversion due to numerical errors). 

Accounting for these correlations, we find that the exozodi of $\beta$~Pictoris\ is best modeled as a compact ($\sim$0.05 au), highly inclined emission near the dust sublimation radius of carbonaceous grains, with a PA misaligned by $\sim 60\degree$ relative to the known cold debris disk. This misalignment {could} point to an exocometary production mechanism for the dust. {The precise radial distribution is not constrained by the data, and both a ring and a 2D Gaussian explain the data equally well}. We find that the flux from this emission represents between 2$\%$ and 7$\%$ of the total flux (depending on the model chosen) at $\lambda = 3.4~\mu$m. We also {note that the data cannot distinguish between the one-component and two-component models with} a more extended emission at $\sim$ 1 au, which emits approximately four times less flux than the inner component. It is {also} important to note that, just as sparse \textit{uv} sampling can affect the retrieval of the flux ratio under the hypothesis of an overextended emission, it can also bias the retrieved values for the PA and inclination of the emission. Further observations with a denser \textit{uv} plane sampling should be performed to further confirm the orientation of the emission found in this study.

Finally, the \textit{L}-band spectral slope suggests the presence of submicron-sized grains. This study offers the first direct observational evidence supporting SED-based predictions for hot exozodis, both in terms of dust location and grain size, while also reinforcing the case for an exocometary origin. The method developed in this study can now be applied to other planetary systems hosting hot exozodis observed with the VLTI / MATISSE instrument. It can also be extended to other instruments such as VLTI / GRAVITY to extend the spectral coverage.

\begin{acknowledgements}
      MATISSE is a consortium composed of institutes in France (J-L Lagrange Laboratory, INSU-CNRS, C\^ote d’Azur Observatory, the University of Nice Sophia-Antipolis), Germany (MPIA, MPIfR, and the University of Kiel), the Netherlands (NOVA and the University of Leiden), and Austria (the University of Vienna). The Konkoly Observatory and the University of Cologne have also provided support in manufacturing the instrument.
      NAOMI corrective optics were developed for ESO by IPAG (VLT-SOW-ESO-15190-5923). 
    We thank Antonio Claret for the help in calculating the limb-darkening coefficients used in this study and Alexandre Gallenne for useful discussion on the data interpretation. PP, JCA and JM acknowledge financial support from the Programme National de Planétologie (PNP) of CNRS-INSU in France, through the EPOPEE project (Etude des POussières Planétaires Et Exoplanétaires). This research has made use of the Jean-Marie Mariotti Center (JMMC) - AMHRA service at \href{https://amhra.jmmc.fr}{https://amhra.jmmc.fr} and has benefited from the help of SUV, the VLTI user support service of the Jean-Marie Mariotti Center (\href{https://www.jmmc.fr/suv}{https://www.jmmc.fr/suv}). D.D. acknowledges support from the European Research Council (ERC) under the European Union's Horizon 2020 research and innovation program (grant agreement No. CoG - 866070). JV is funded from the Hungarian NKFIH OTKA project no. K-132406, and this work was also supported by the NKFIH NKKP grant ADVANCED 149943. Project no.149943 has been implemented with the support provided by the Ministry of Culture and Innovation of Hungary from the National Research, Development and Innovation Fund, financed under the NKKP ADVANCED funding scheme. TAS acknowledges financial support from the National Aeronautics and Space Administration (NASA) through grants 80NSSC23K1473 and 80NSSC23K0288. { We thank the referees for their constructive comments and suggestions, which helped to significantly strengthen and improve the clarity of this manuscript.
}
\end{acknowledgements}

\bibliographystyle{aa}  
\bibliography{bpic} 
%

\begin{appendix} 

\section{$\beta$~Pictoris parameters}

{The key parameters of $\beta$~Pictoris and its two known planets are summarized in Table~\ref{tab:bpicproperties}}.

\begin{table}[h]
    \centering
    \caption{Parameters for the star, outer debris disk, and two known planets.}
    \begin{tabular}{cccc}
    
        \hline
        \hline
        
        \vspacetab & Parameter & Value & References\\
          
        \hline
        
        \multirow{1}{5em}{\textbf{Star}} &\vspacetab $T_{\textrm{eff}}$ & $7938 ^{+5.8}_{-6.1}$ K & (1)\\
        \vspacetab & distance & $19.63^{+0.06}_{-0.06}$ pc & (1) \\
        \vspacetab &age & $\sim$ 20 Myr & (2, 3, 4) \\
        \vspacetab &$\theta_{\rm LD}$ & $0.736 \pm 0.019$ mas & (5) \\

        \hline
        
        \multirow{2}{5em}{\textbf{Outer disk}} & \vspacetab  PA\tablefootmark{(a)} & $29.6\degree \pm 1.4\degree$ & (6) \\
        \vspacetab & $i_{\textrm{tilt}}$\tablefootmark{(b)} & $86\degree\pm 1\degree$ & (7, 8, 9) \\

        \hline
        
        \multirow{2}{5em}{\textbf{$\beta$~Pictoris\ b}} & \vspacetab  $a_b$ & $9.93\pm 0.03$~au & (10) \\
        \vspacetab & $i_b$\tablefootmark{(b)} & $89\degree\pm 0.01\degree$ & (10) \\
        \vspacetab & $e_b$ & $0.103\pm 0.003$ & (10) \\
        
        \hline
        
        \multirow{2}{5em}{\textbf{$\beta$~Pictoris\ c}} & \vspacetab  $a_c$ & $2.68\pm 0.02$~au & (10) \\
        \vspacetab & $i_c$\tablefootmark{(b)} & $88.95\degree\pm 0.10\degree$ & (11) \\
        \vspacetab & $e_c$ & $0.32\pm 0.02$ & (10) \\
        
        \hline
        
    \end{tabular}
    \tablebib{
    (1) \citet{gaiaDR3}; 
    (2) \citet{Shkolnik2017};
    (3) \citet{Betapic_age};
    (4) \citet{Couture2023}; 
    (5) \cite{Defrere_2012_betapic};
    (6) \citet{Rebollido2024};
    (7) \citet{Milli2014} ;
    (8) \citet{Millar-blanchaer2015};
    (9) \citet{Matra2019};
    (10) \cite{Lacour2021} }
    \tablefoot{\tablefoottext{a}{The parameters measured here are from the JWST/NIRCam long-wavelength channel ($2.4$--$5~\mu$m). The PA is defined from north to east;} \tablefoottext{b}{the inclination is defined as 0\degree\ for face-on orientation and 90\degree\ for edge-on orientation.}}
    \label{tab:bpicproperties}
\end{table}

\section{MATISSE \textit{L}-band data}\label{sec:closure_phase}

\renewcommand{\arraystretch}{2}
\begin{table*}[h!]
    \caption{VLTI/MATISSE observations of $\beta$~Pictoris.  \label{tab:observations}} 
\resizebox{\textwidth}{!}{\begin{tabular}{cccccccccc}
\hline
\hline
  \textbf{Array} & \textbf{Stations} & \textbf{Date} & \textbf{Chop?} & \textbf{DIT {[}s{]}} & \textbf{Seeing {[}"{]}} & \textbf{$\tau_0$ {[}ms{]}} & \textbf{IWV {[}mm{]}} & \textbf{Sequence} & \textbf{Calibrator} \\ 
\cline{1-10}
 UTs   & U1-U2-U3-U4 & \multicolumn{1}{c}{2021-01-08}   & No & 0.111 & \multicolumn{1}{c}{0.85}  & \multicolumn{1}{c}{6.0}& \multicolumn{1}{c}{4.5}  & CAL-SCI   & HD~28413   \\ 
\cline{1-10}
 {Large}   & A0-G1-J2-J3 & \multicolumn{1}{c}{2021-01-16}& No  &0.111& \multicolumn{1}{c}{0.55}  & \multicolumn{1}{c}{5.6}& \multicolumn{1}{c}{1.6}  & CAL-SCI  & HD~33042   \\ 
\cline{1-10}
 Small & A0-B2-D0-C1 & \multicolumn{1}{c}{2021-01-28}& Yes &0.111& \multicolumn{1}{c}{0.555} & \multicolumn{1}{c}{8.3} & \multicolumn{1}{c}{4.6}  & CAL-SCI  & HD~33042 \\ 
\cline{1-10}
 Small   & A0-B2-D0-C1& \multicolumn{1}{c}{2022-10-14}& Yes & 0.6   & \multicolumn{1}{c}{0.67}  & \multicolumn{1}{c}{7.5}& \multicolumn{1}{c}{1.0}  & CAL1-SCI-CAL2-SCI  & HD~40091-HD~40808   \\ \cline{1-10}
\end{tabular}}
\tablefoot{
 The fourth column indicates whether chopping was used. The last two columns indicate the calibrating sequence and the calibrators used. The acronym IWV stands for integrated water vapor. The calibrator angular sizes are taken from the JSDC catalog \citep{JSDC}}
\end{table*}

\begin{figure*}[h]
    \centering
    \includegraphics[width=0.9\textwidth]{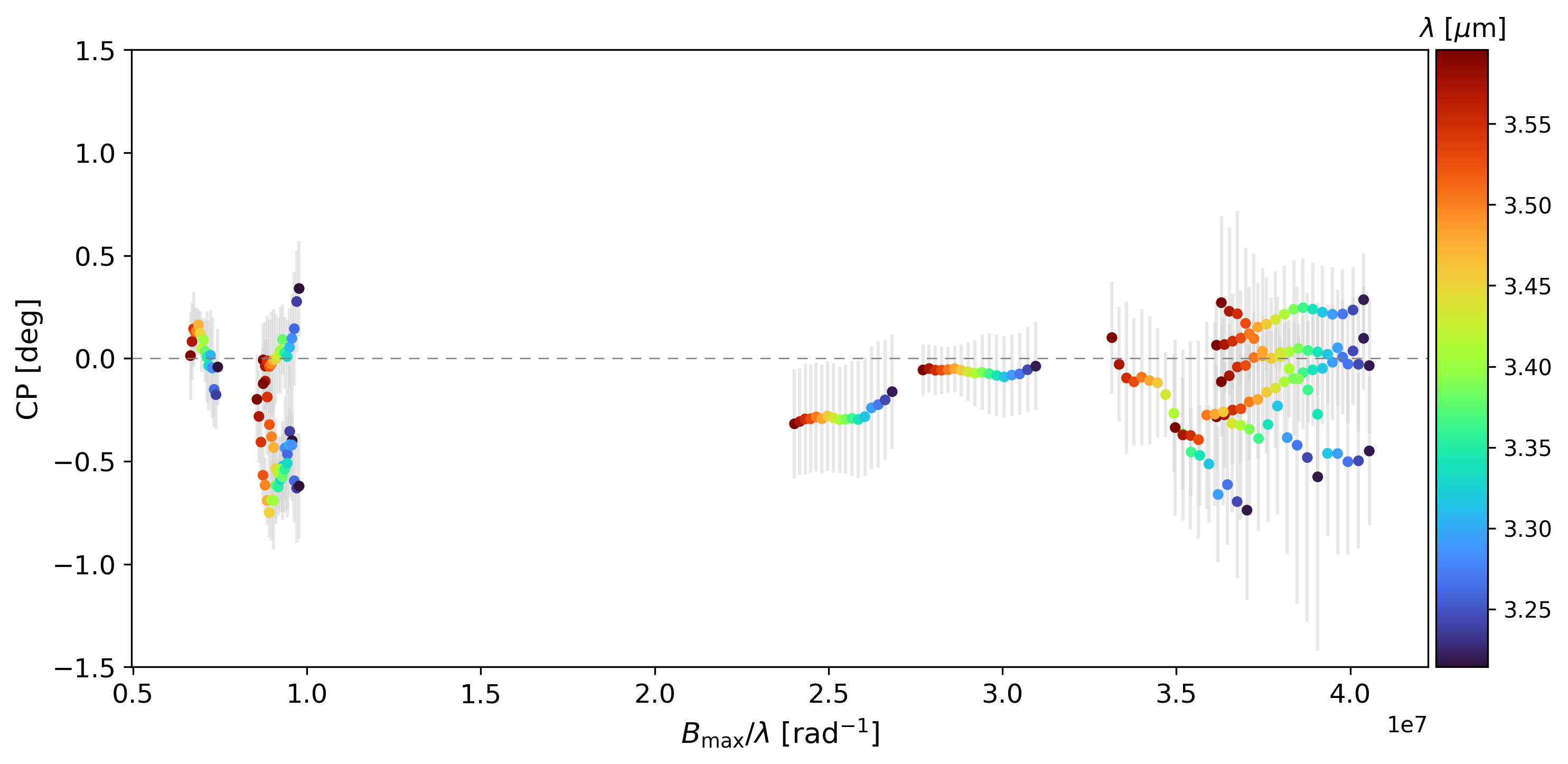}
    \caption{\textit{L}-band interferometric CPs as a function of the spatial frequency of the longest baseline in the triangle.}
    \label{fig:CP}
\end{figure*}

The closure phases for the \textit{L}-band MATISSE observations are shown in Fig.~\ref{fig:CP}. They are generally close to zero, although a weak signal may be present in the AT small 2021 dataset. While these closure phases could be used to place upper limits on the presence of a companion, such an analysis is beyond the scope of this study.

\section{MATISSE \textit{L}-band data: Correlations in the intermediate frames during an exposure}
\label{sec:correlations_raw_data}

We present the correlations between spectral channels, baselines, and telescopes in both the photometry and {correlated flux} intermediate frames, obtained during a single exposure with the MATISSE instrument in the \textit{L} band (i.e., before time-averaging; see Sect.~\ref{sec:final_step_of_data_reduction} for details). The intermediate data show strong correlations, which likely arise from a variety of complex instrumental and observational effects. This suggests that the final data products will also exhibit significant correlations, which we estimate using an empirical approach.

\subsection{Correlations between wavelengths for the photometry and correlated flux}
\label{sec:correlations_wavelengths}
We study the correlations between different spectral channels for both the photometry and the {correlated flux}. To calculate this correlation on the photometric frames, we first sum the values for the pixels along the spatial direction of the detector and then calculate the correlations between spectral channels. We repeat this for each telescope and each BCD position. For the {correlated flux}, we first calculate the modulus $\left | \underline{\textbf{I}}(u, \lambda, t)\right|$, then integrate the values of the resulting frame between $(B_{ij} - D)/\lambda$ and $(B_{ij} + D)/\lambda$ for each of the six baselines. Finally we calculate the correlations of this quantity between spectral channels for each baseline. We repeat this step for each BCD position.

We find that the spectral channels of both frame types are strongly correlated or anticorrelated between each other (the correlation coefficient is very close to 1 or -1). This same effect was observed for GRAVITY data \citep{GRAVITY_correlations}, where the origin of these correlations is believed to be a combination of the shared optical path for all spectral channels and the interpolation of pixels to form the spectral channels (because one pixel does not correspond to a spectral channel). For MATISSE data we also believe that these are the predominant effects.  

\subsection{Correlations between telescopes for the photometry}

\begin{figure}[tbp!]
    \centering
    \includegraphics[width=0.5\textwidth]{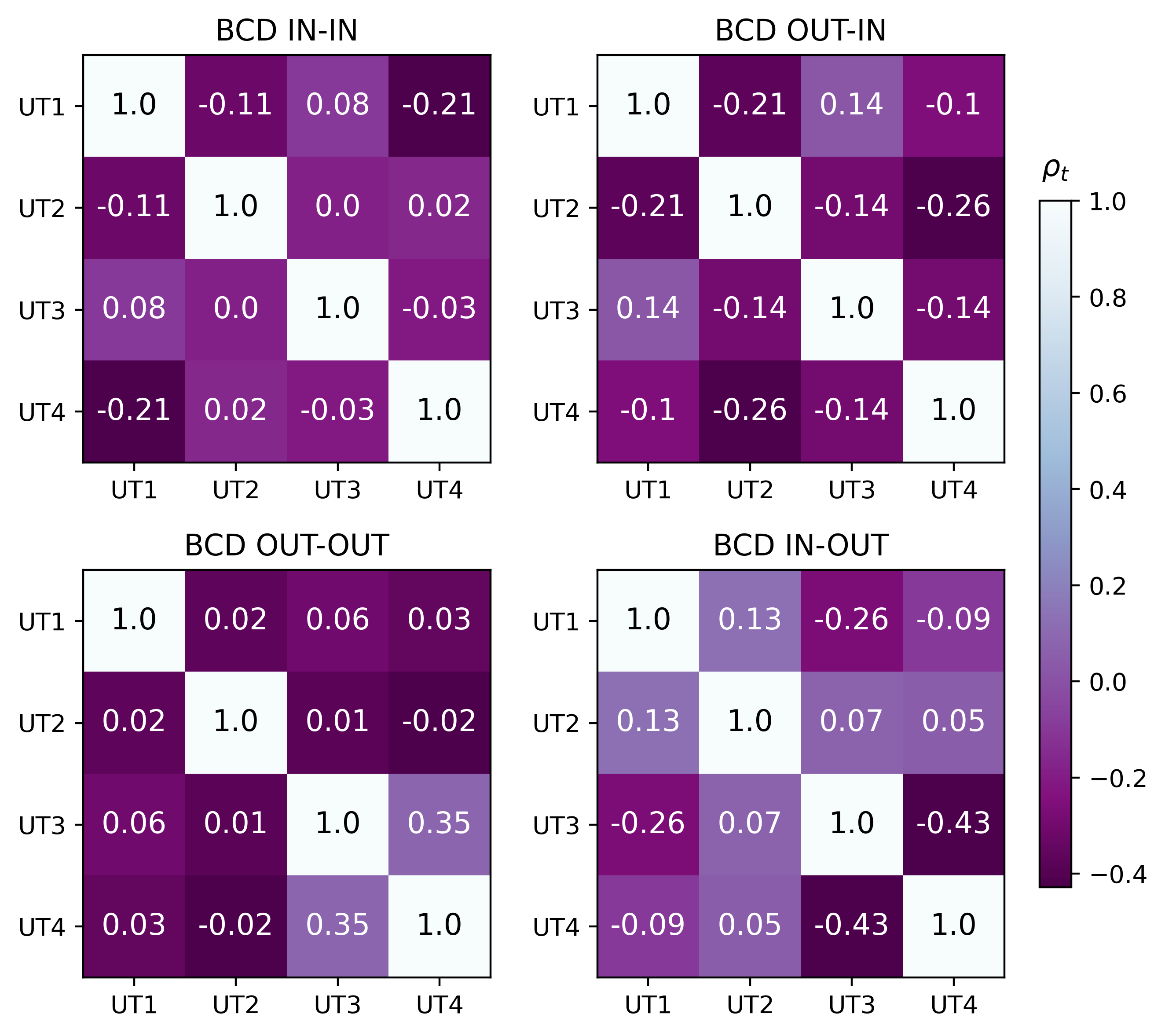}
    \caption{Correlation coefficients ($\rho_t$) between telescopes for the photometry of the UT dataset on the science target ($\beta$~Pictoris) at each BCD position.}
    \label{fig:corr_tels}
\end{figure}

We calculate the correlation between the photometric measurements of each telescope for each BCD position. Given that the spectral channels are strongly correlated (see Sect.~\ref{sec:correlations_wavelengths}), we choose $\lambda=$ 3.5 $\mu$m (where the transmission is higher in the \textit{L} band) as a reference wavelength to study how the measurements of the photometry for different telescopes are correlated. We sum the values for the pixels along the spatial direction for this spectral channel and compute the correlation coefficient of this quantity between telescopes. The results for the UT array observations are presented in Fig. \ref{fig:corr_tels}.

\subsection{Correlations between baselines for the correlated flux}
\begin{figure}[tbp!]
    \centering
    \includegraphics[width=0.5\textwidth]{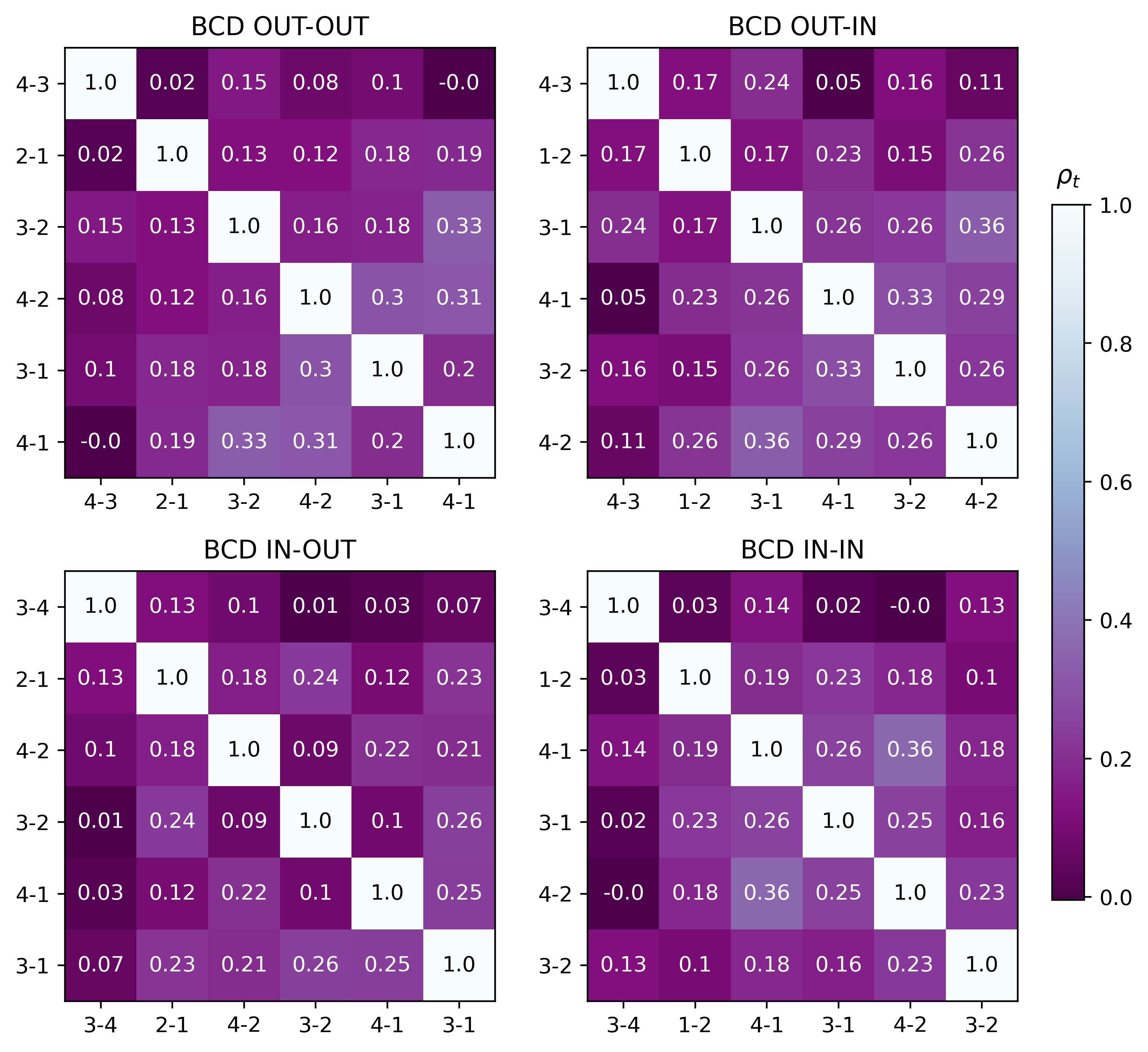}
    \caption{Correlation coefficients ($\rho_t)$ between baselines for $\left | \underline{\textbf{I}}(u, \lambda, t)\right|$ of the AT Large Medium dataset on the science target, for each BCD position. The numbers represent the telescope indices (e.g., the baseline formed by AT3 and AT2 is labeled 3-2}).
    \label{fig:fig_corr_base}
\end{figure}

To study the correlations of the {correlated flux} between baselines, we first calculate the modulus $\left | \underline{\textbf{I}}(u, \lambda, t)\right|$, then integrate the resulting frame between $(B_{ij} - D)/\lambda$ and $(B_{ij} + D)/\lambda$ for each of the six baselines. As for the photometry, since all spectral channels are correlated (Sect.~\ref{sec:correlations_wavelengths}), we evaluate the quantity at $\lambda=$ 3.5 $\mu$m. The results for the AT Large Medium array observations are presented in Fig. \ref{fig:fig_corr_base}.

On average, the correlation of two baselines sharing a telescope is higher than that of independent baselines. However, this is not the only source of correlation between baselines. In MATISSE, the beams are configured in a nonredundant way so that the fringe peaks are separated in Fourier space. However, this induces another source of correlation, where baselines projected closely in Fourier space are more correlated together. This proximity changes with each BCD configuration. 

\section{{Principal component analysis based method}}
\subsection{Distribution along the principal components} \label{sec:distribution_along_PC}

{An important assumption in applying the approach described in Sect. \ref{sec:correlations_general_approach} is that the frames are not correlated with each other (i.e., there is no time correlation). We verify this assumption in this section.}

\begin{figure}[h]
    \centering
    \includegraphics[width=0.4\textwidth]{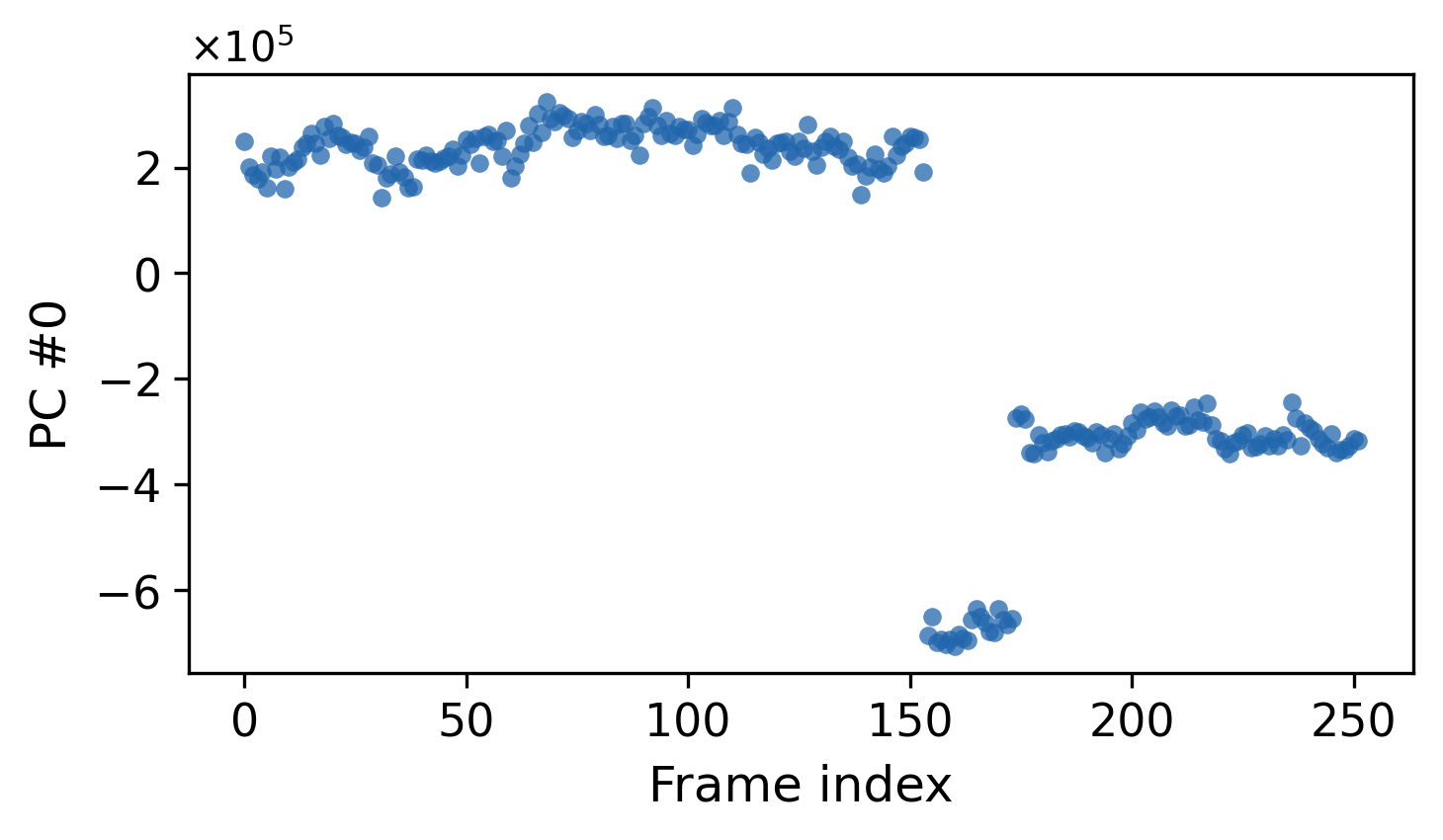}
    \caption{Example data distribution along the first PC (here for the BCD position IN-IN), showing a strong multimodal behavior.}
    \label{fig:non_gauss_PCA}
\end{figure}

Indeed, when the features of the PCA contain both the imaginary and real parts of the {correlated flux}, the distribution of the data along the PCs is non-Gaussian, often multimodal (an example is shown in Fig. \ref{fig:non_gauss_PCA} for the calibrator of the Large Medium observations with BCD position IN-IN) {and most importantly, the values along the PC are strongly time-dependent, making them time correlated}. This effect appears to be caused by significant phase jumps in the fringes, leading to large non-Gaussian variations in the real and imaginary parts of the {correlated flux}, even though the absolute value remains stable during these events. For this reason, in Sect. \ref{sec:PCA} we chose to use the absolute value of the {correlated flux} as part of the features. Since we are studying the correlations in the visibility, any phase effects should not impact the results and therefore this approach is justified.

When we use the absolute value, instead of the real and imaginary parts of the {correlated flux}, we find that the values along the PCs are uncorrelated with time. This allows us to resample the points independently (as detailed in Sect. \ref{sec:PCA}) to create a resampled dataset. We then apply the inverse transform to this dataset to obtain our synthetic MATISSE intermediate frames $\left | \underline{\textbf{I}}(u, \lambda, t)\right|$ and $P_{i}(x, \lambda, t)$. Finally we process these frames with the standard reduction pipeline to obtain squared visibilities. We repeat this for the science target and the calibrator(s) independently, and calibrate the data accordingly.

\subsection{{Preservation of information with a PCA approach}}
\label{sec:preservation_of_correlations}

\begin{figure}[tbph!]
    \centering
    \includegraphics[width=0.4\textwidth]{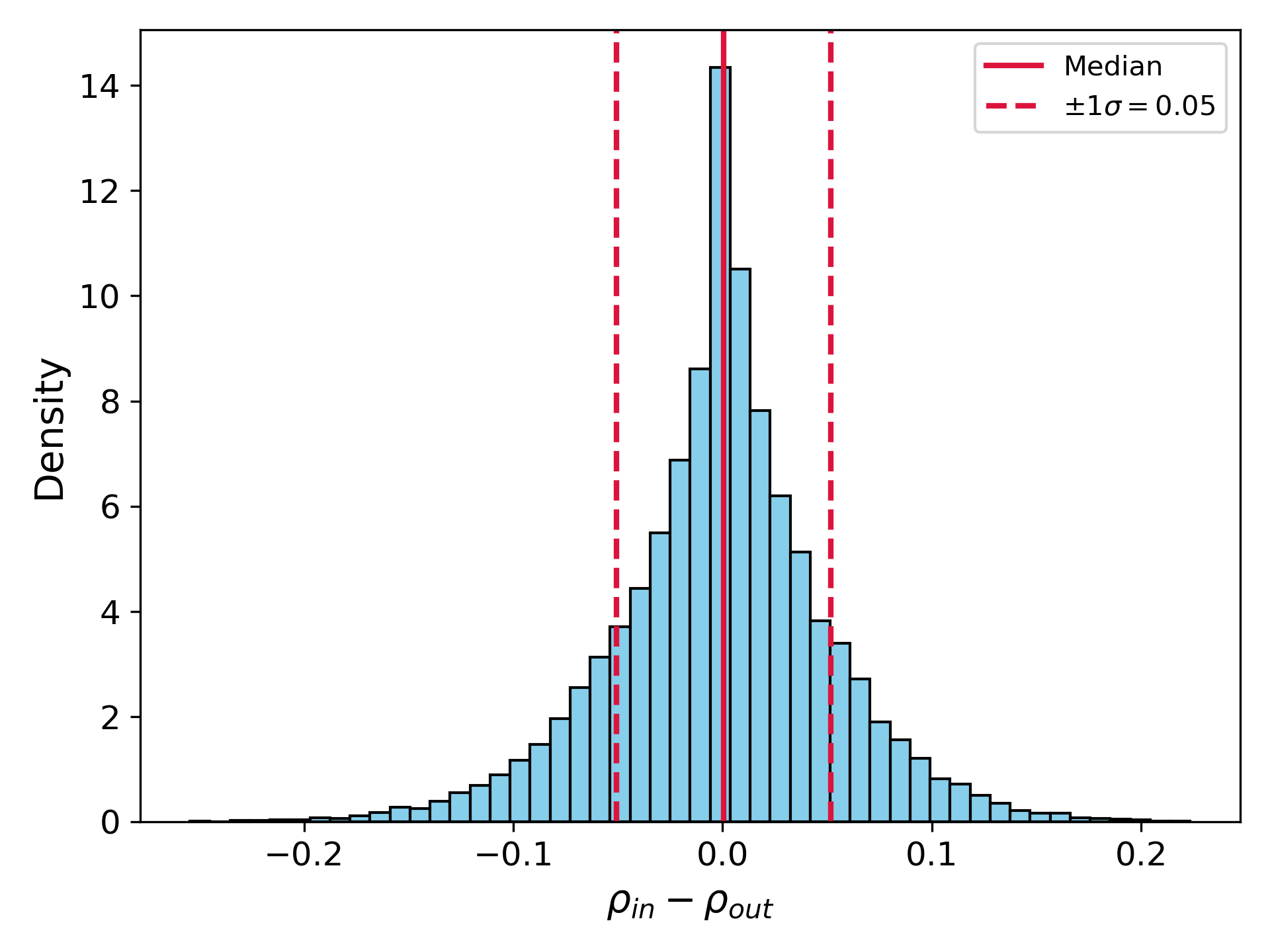}
    \caption{{Distribution of the difference between the retrieved and true values for a correlation in visibility space across noise iterations and correlation values.}}
    \label{fig:corr_retrieval_results}
\end{figure}

{{To ensure correlations are preserved with our method, we have generated synthetic data with known correlations and compare them with the correlations obtained with our approach. We start by generating a dataset with 4 elements containing data and $N_{noise}$ elements containing noise ($N_{noise}\sim30$). Two of the data elements represent correlated flux frames, the other two represent photometric frames. We impose a certain correlation between the two pixels for each type of frame, and a cross-correlation between the two types of frames. We assume the noise is uncorrelated. We then calculate a "visibility" as the ratio between the correlated flux elements and the photometric elements. We measure the correlation between the two visibility elements. We then apply our PCA method to this dataset and generate with it a synthetic dataset. With this dataset we calculate a visibility and compute its correlations. We repeat this process a number of times where we vary the correlation strength and the noise realization across iterations. In Fig.~\ref{fig:corr_retrieval_results}, the difference between the input correlation and the retrieved correlation is shown. Our approach preserves correlations in the visibility measurement using the PCA approach and that the standard deviation of the difference across iterations is 0.05. }
}

\section{Analytical expressions for the visibility of the geometric models} 
\label{sec:expressions_for_V2}
In all these expressions, $u$ and $v$ represent the spatial frequencies sampled, with $u=B_u/\lambda$ and $v=B_v/\lambda$, where $B_v$ is the projected baseline along the south-north direction, $B_u$ the projected baseline along the east-west direction and $\lambda$ the wavelength of observation. For all expressions we use common notations:
\begin{eqnarray}
    B/\lambda & = & \sqrt{u^2+v^2}, \\
    \rho & = & \sqrt{u_{\textrm{eff}}^2 + v_{\textrm{eff}}^2},
\end{eqnarray}
where
\begin{eqnarray}
    u_{\textrm{eff}} & = & \left(u\cos(\textrm{PA}) - v\sin(\textrm{PA}) \right)\times \cos(i_{\textrm{tilt}}) \\
    v_{\textrm{eff}} & = & u\sin(\textrm{PA}) + v\cos(\textrm{PA})
\end{eqnarray}
are the effective spatial frequencies in the circumstellar component's reference frame after correction for the inclination ($i_{\mathrm{tilt}}$) and PA of the apparent major axis. The $u_{\textrm{eff}}$ and $v_{\textrm{eff}}$ spatial frequencies define the sampling points for evaluating the geometric models of the circumstellar component presented below

For the 2D Gaussian, we also provide the relation between the FWHM of the Gaussian  and the half-light radius ($r_{\textrm{HL}}$), defined as the radius enclosing half of the total flux. This quantity is useful for comparing spatial extents across models. For the Gaussian ring, the expression for the half-light radius cannot be solved analytically. Therefore, in the histograms in Sects.~\ref{sec:Models} and \ref{sec:postdist_with_boot}, we present the half-light radius for the 2D Gaussian and the radius of the Gaussian ring. 
{

All the geometric models for the circumstellar emission assume that the emission has negligible vertical extent. If this assumption is untrue it would manifest as a smaller inferred inclination than the true one. Therefore, derived inclinations should be interpreted as lower limits if the disk thickness is significant.
}
\subsection{Stellar model}
The expression for the visibility of the limb-darkened photosphere is \citep{limbdarkenedphotosphere}
\begin{equation}
    V_{\star}(B/\lambda) = \frac{6}{(3 - u_{\lambda})}\left[(1-u_{\lambda}) \frac{J_1(x)}{x} + u_{\lambda} 
    \sqrt{\frac{\pi}{2}}\frac{J_{3/2}(x)}{x^{3/2}}\right],
    \label{eq:Vstellar}
\end{equation}
where $u_\lambda$ is the linear limb-darkening coefficient, $x = \pi  \theta_{\rm LD} B/\lambda$, $\theta_{\rm LD}$ is the limb-darkened angular diameter of the star (for the \textit{L} band in this case) and $J_1(x)$ and $J_{3/2}(x)$ are the Bessel functions of order 1 and 3/2, respectively.

\subsection{2D Gaussian}
The expression for the visibility, $V(u,v)$, of the 2D Gaussian radial profile model is as follows:
\begin{equation}
    V(u,v) = \exp{\left(-\frac{(\pi \theta \rho)^2}{4\ln{2}}\right)}
    \label{eq:V2Dgaussian},
\end{equation}
where $\theta$ is the FWHM of the Gaussian. The half-light radius is
\begin{equation}
    r_{\textrm{HL}} = \frac{\theta}{2} .
\end{equation}
In Sects.~\ref{sec:spatial_info} and \ref{sec:twocomponentmdls}, the parameter $\theta$ in this model is denoted as the FWHM.
\subsection{Gaussian ring}
The expression for the visibility, $V(u,v)$, of the ring model with a Gaussian radial profile is as follows:
\begin{equation}
    V(u,v) = \exp\left({-\frac{\left(\pi \theta \rho \right)^2}{4\ln{2}}}\right) \times J_0\left(2\pi r_{\text{ring}} \rho \right),
    \label{eq:Vgaussianring}
\end{equation}
where $\theta$ is the FWHM of the Gaussian (which translates the width of the ring) and $r_{\text{ring}}$ is the radius of the ring. In Sects.~\ref{sec:spatial_info} and \ref{sec:twocomponentmdls}, the parameter $\theta$ in this model is referred to as $\textrm{w}_{\textrm{FWHM}}$.

\subsection{Spectral models}
To incorporate wavelength dependence into the flux ratio, we use the following expression:
\begin{equation}
    f_{\lambda} = f_{3.4\, \mu \mathrm{m}} \times \left(\frac{\lambda_{\mu \mathrm{m}}}{3.4\, \mu \mathrm{m}}\right)^{p_{\mathrm{spec}}},
    \label{eq:spectralslope}
\end{equation}
where $f_{3.4\, \mu \mathrm{m}}$ and $p_{\mathrm{spec}}$ are two free parameters.

\section{{Likelihoods}}
\label{sec:appdx_logl}
{

In the Bayesian inference framework, correlations between different data points are taken into account through the likelihood function, $\mathcal{L}(\theta)$, expressed as
\begin{equation}\label{eq:LLcorr}
     -2\textrm{ln} \mathcal{L}(\theta) \equiv \chi^2 = (\textbf{d}-\textbf{m}(\theta))^T \mathbf{\Sigma}^{-1}(\textbf{d}-\textbf{m}(\theta)),
\end{equation}
where $\theta$ represents the model parameters, $\mathbf{d}$ is the data vector, $\mathbf{m}(\theta)$ is the model vector, and $\mathbf{\Sigma}^{-1}$ is the inverse covariance matrix of the data (also called precision matrix). When assuming that data points are independent (i.e., $\Sigma$ is a diagonal matrix with entries $\sigma_i^2$ on the main diagonal, 
where $\sigma_i$ is the variance of data point $d_i$), Eq.~\ref{eq:LLcorr} simplifies to $-2\mathrm{ln} \mathcal{L}(\theta) \equiv \chi^2 =  \sum_i \left( \frac{d_i-m_i(\theta)}{\sigma_i}\right)^2$.

In practice, since the covariance matrix is not known a priori but rather estimated from the data, it can be best represented as a random object with some intrinsic uncertainty. Therefore, following the results of \citet{LL_for_estimated_cov_mtrx} we use the following non-Gaussian expression for the likelihood:

\begin{equation}
\label{eq:new_LL}
\mathcal{L}(\theta) = \bar{c}_{\rm p} |{\bf{\Sigma}}|^{-1/2} 
\left[ 1 + \frac{(\textbf{d}-\boldsymbol {\textbf{m}(\theta) })^T{\bf{\Sigma}}^{-1} (\textbf{d}- \boldsymbol {\textbf{m}(\theta) }) }{N-1}\right] ^{\frac{-N}{2}},
\end{equation}
where $\mathbf{\Sigma}$ is the estimated covariance matrix, $N$ is the number of samples used in the estimate, and
\begin{equation}
\bar{c}_{\rm p} = \frac{\Gamma \left( \frac{N}{2} \right)}{ \left[\pi (N-1)\right]^{p/2} \Gamma \left( \frac{N-p}{2} \right) },
\end{equation}
with $\Gamma$ the Gamma function and $p$ the dimension of the vector \textbf{d}.

In Tables \ref{ORDresults}, \ref{tab:models_one_component} and \ref{twocompresults} we use equation \ref{eq:LLcorr} to calculate the reduced $\chi^2$ ($\chi^2_r = \frac{\chi^2}{N_{data}-N_{params}}$ where $N_{data}$ is the number of data points and $N_{params}$ is the number of free parameters for the given model). 

To calculate the Akaike information criterion (AIC), we used the following expression:
\begin{equation}
\label{eq:AIC}
    AIC = 2k + \chi^2,
\end{equation}
where $k$ is the number of free parameters for the model in question.
The AIC provides an additional criterion for model selection. Models are compared based on their AIC values, with smaller AIC indicating a better trade-off between goodness of fit and model complexity, and therefore a preferred model.

}

{

In this study we use Nested Sampling instead of Markov Chain Monte Carlo (MCMC) because our posterior distributions contain local minima, requiring a large number of iterations for the walkers to converge on the global minimum, whereas the Nested Sampling algorithm appeared to converge more efficiently to the global minimum. Another advantage of this method is that it directly estimates the Bayesian evidence, which we use to compare the goodness of fit for models with different degrees of freedom.

These algorithms (MCMC and Nested Sampling) are also used to estimate the errors on the best-fit parameters. However, in our case, these errors appear unrealistically small when accounting for correlations. Therefore, to estimate the errors on the best-fit parameters, we use nonparametric bootstrapping. This involves generating new synthetic datasets -- using the covariance matrices calculated in Sect.~\ref{sec:correlations} and the real data as the median -- and fitting the models to these datasets. We repeat this process 1000 times and define the variance of the best-fit parameters across iterations as the errors. Notably, since the uncertainties are more conservative than those estimated with Nested-Sampling alone we report the former in the tables.
}
\section{Priors used for nested sampling}
\label{sec:appdx_prior}
For all parameters we use a uniform prior $\mathcal{U}(a,b)$. See Tables \ref{tab:uniform_priors_singlecomp} and \ref{tab:twocomp_priors} for the bounds on the prior of each parameter. 
\begin{table*}[h]
\centering
\caption{Uniform prior ranges adopted for the single-component emission models.}
\label{tab:uniform_priors_singlecomp}
\footnotesize
\setlength{\tabcolsep}{2.8pt}
\renewcommand{\arraystretch}{1.12}

\begin{tabular}{lccccccc}
\hline
\hline
Model 
& FWHM [mas]
& $r_{\rm ring}$ [mas] 
& $w_{\rm FWHM}$ [mas]
& $\textrm{cos}(i_{\rm tilt})$ 
& PA [deg]
& $f$ [$\%$]
& $p_{\rm spec}$ 
 \\
\hline

ORD 
& — 
& — 
& — 
& —  
& —  
& $\mathcal{U}(0,\,100)$ 
& —  \\

ORDS 
& — 
& — 
& — 
& —  
& —  
& $\mathcal{U}(0,\,100)$ 
& $\mathcal{U}(-10,\,10)$ 
 \\
 
G 
& $\mathcal{U}(0,\,400)$ 
& — 
& — 
& $\mathcal{U}(0,\,1)$ 
& $\mathcal{U}(-90,\,90)$ 
& $\mathcal{U}(0,\,100)$ 
& —  \\

GS 
& $\mathcal{U}(0,\,400)$ 
& — 
& — 
& $\mathcal{U}(0,\,1)$ 
& $\mathcal{U}(-90,\,90)$ 
& $\mathcal{U}(0,\,100)$ 
& $\mathcal{U}(-10,\,10)$ 
 \\

GR 
& — 
& $\mathcal{U}(0,\,200)$ 
& $\mathcal{U}(0,\,100)$ 
& $\mathcal{U}(0,\,1)$ 
& $\mathcal{U}(-90,\,90)$ 
& $\mathcal{U}(0,\,100)$ 
& — 
 \\

GRS 
& — 
& $\mathcal{U}(0,\,200)$ 
& $\mathcal{U}(0,\,100)$ 
& $\mathcal{U}(0,\,1)$ 
& $\mathcal{U}(-90,\,90)$ 
& $\mathcal{U}(0,\,100)$ 
& $\mathcal{U}(-10,\,10)$ 
\\

\hline
\hline
\end{tabular}

\end{table*}

\begin{table*}[h]
\centering
\caption{Uniform prior ranges adopted for the two-component emission models.}
\label{tab:twocomp_priors}
\footnotesize
\setlength{\tabcolsep}{2.6pt}
\renewcommand{\arraystretch}{1.12}

\begin{tabular}{lccccccccc}
\hline
\hline
Model
& FWHM$_{\rm inner}$ [mas]
& $r_{\rm inner}$ [mas]
& $r_{\rm outer}$ [mas]
& $w_{\rm FWHM, outer}$ [mas]
& $\textrm{cos}(i_{\rm tilt})$
& PA [deg]
& $f_{\rm inner}$ [$\%$]
& $f_{\rm outer}$ [$\%$] \\
\hline

GIRO
& $\mathcal{U}(0,\,20)$
& —
& $\mathcal{U}(20,\,200)$
& $\mathcal{U}(0,\,50)$
& $\mathcal{U}(0,\,1)$
& $\mathcal{U}(-90,\,90)$
& $\mathcal{U}(0,\,100)$
& $\mathcal{U}(0,\,100)$ \\

RIRO
& —
& $\mathcal{U}(0,\,30)$
& $\mathcal{U}(20,\,200)$
& $\mathcal{U}(0,\,50)$
& $\mathcal{U}(0,\,1)$
& $\mathcal{U}(-90,\,90)$
& $\mathcal{U}(0,\,100)$
& $\mathcal{U}(0,\,100)$ \\

\hline
\hline
\end{tabular}
\end{table*}

\FloatBarrier

\section{Additional posterior distributions}\label{sec:postdist_with_boot}

\begin{figure}[h!]
    \centering
    \includegraphics[width=0.45\textwidth]{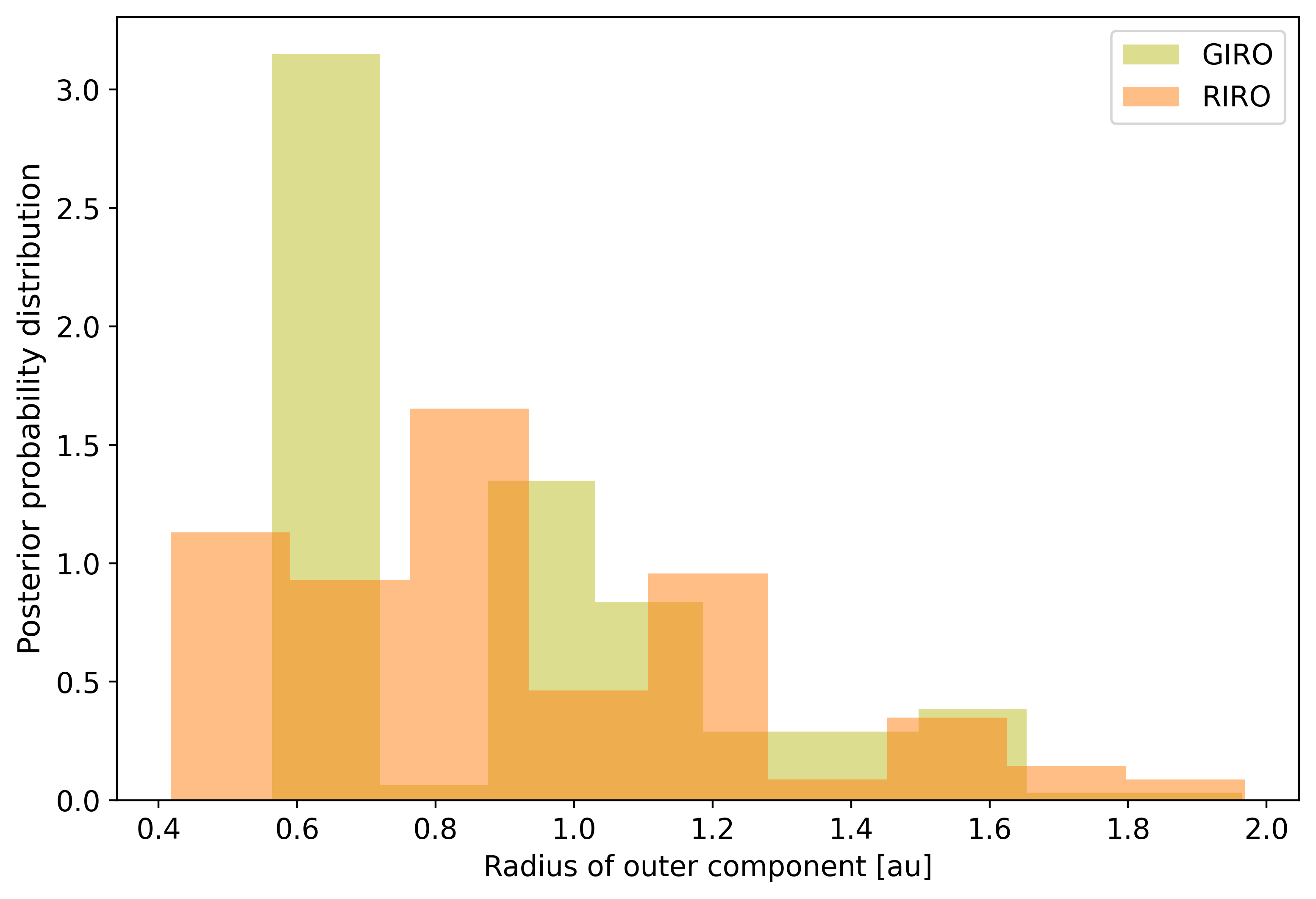}
    \caption{Half-light radius of the outer component of the two-component models.}
    \label{fig:post_Rhl_outer}
\end{figure}

\begin{figure}[h!]
    \centering
    \includegraphics[width=0.45\textwidth]{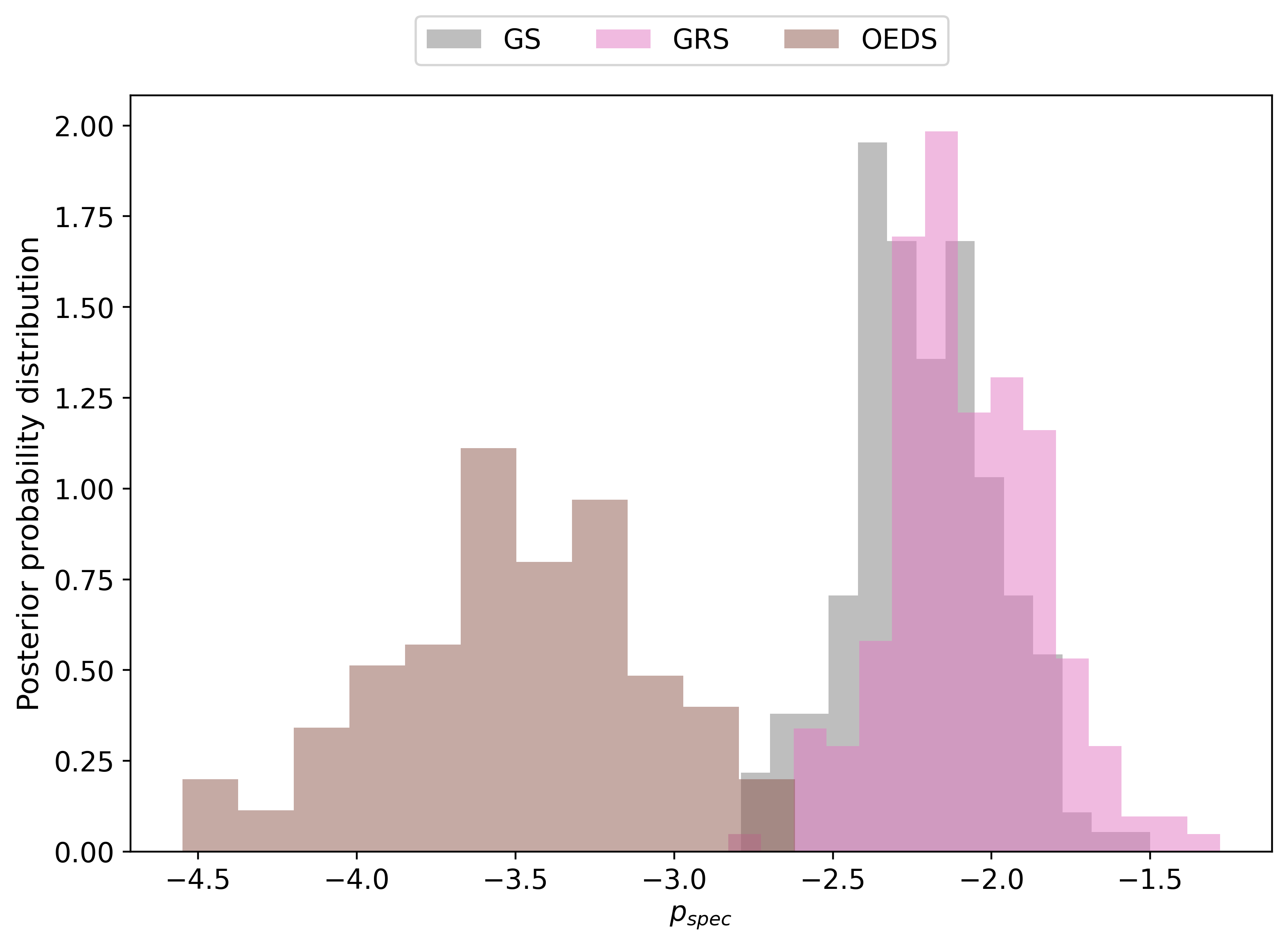}
    \caption{Spectral index of the flux ratio.}
    \label{fig:posterior_spectral_index}
\end{figure}

\begin{figure*}[tbph!]
    \centering
    \includegraphics[width=0.99\textwidth]{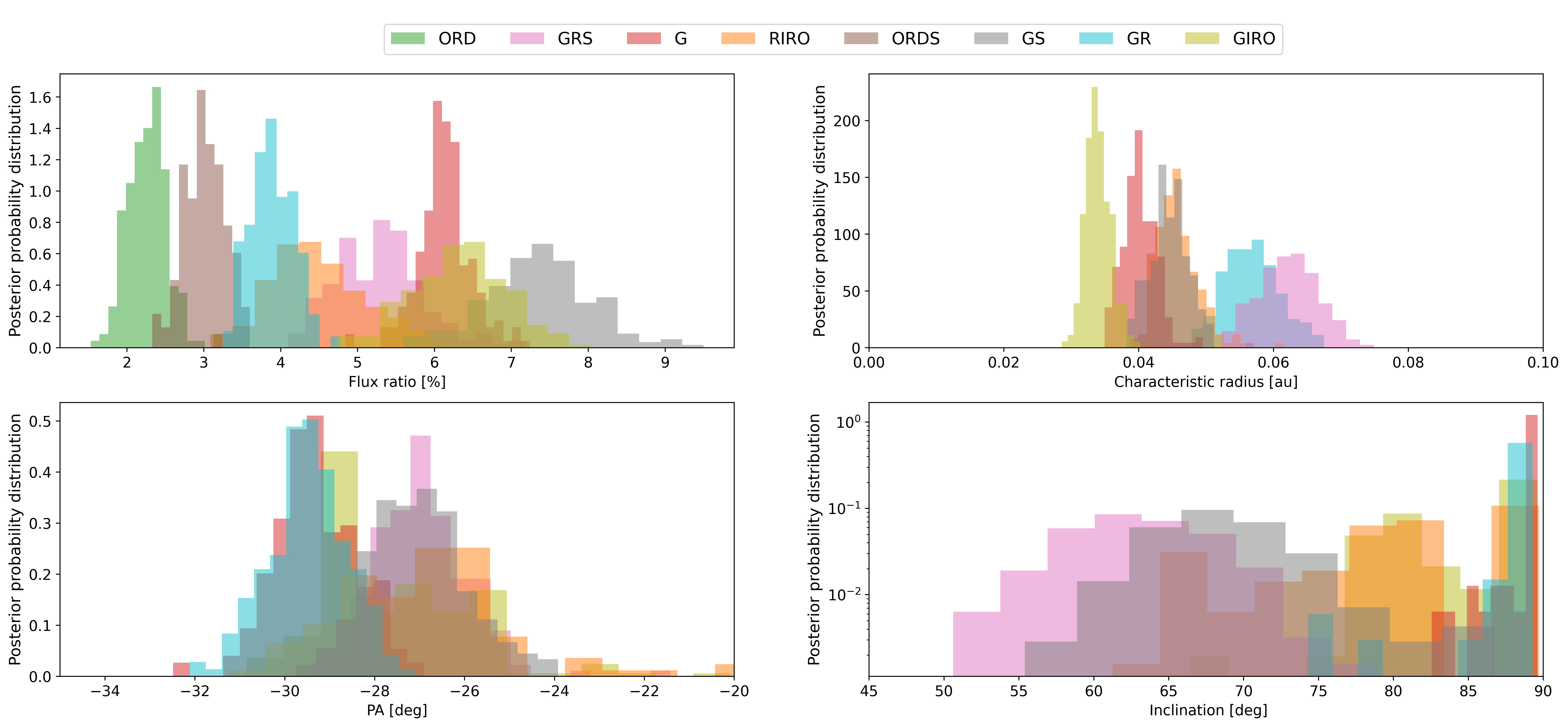}
    \caption{Posterior distributions for all single-component models and the inner component of two-component models, obtained when accounting for correlations. The distribution of inclinations (bottom right) is presented in log scale, because some (e.g., GRS and GS) are broader than others (e.g., G), while the other distributions are shown in linear scale.}
    \label{fig:posterior_dists_incomp_onecomp}
\end{figure*}

Figures~\ref{fig:post_Rhl_outer} and \ref{fig:posterior_spectral_index} show the posterior distributions for the half-light radius of the outer component of the two-component models and spectral index of the flux ratio, respectively, complementing those in Fig.~\ref{fig:posterior_dists_incomp_onecomp}.

\section{Best fits to the \textit{L}-band MATISSE visibilities}\label{sec:appdx_best_fits}

\begin{figure*}[h!]
    \centering
    \includegraphics[width=1\textwidth]{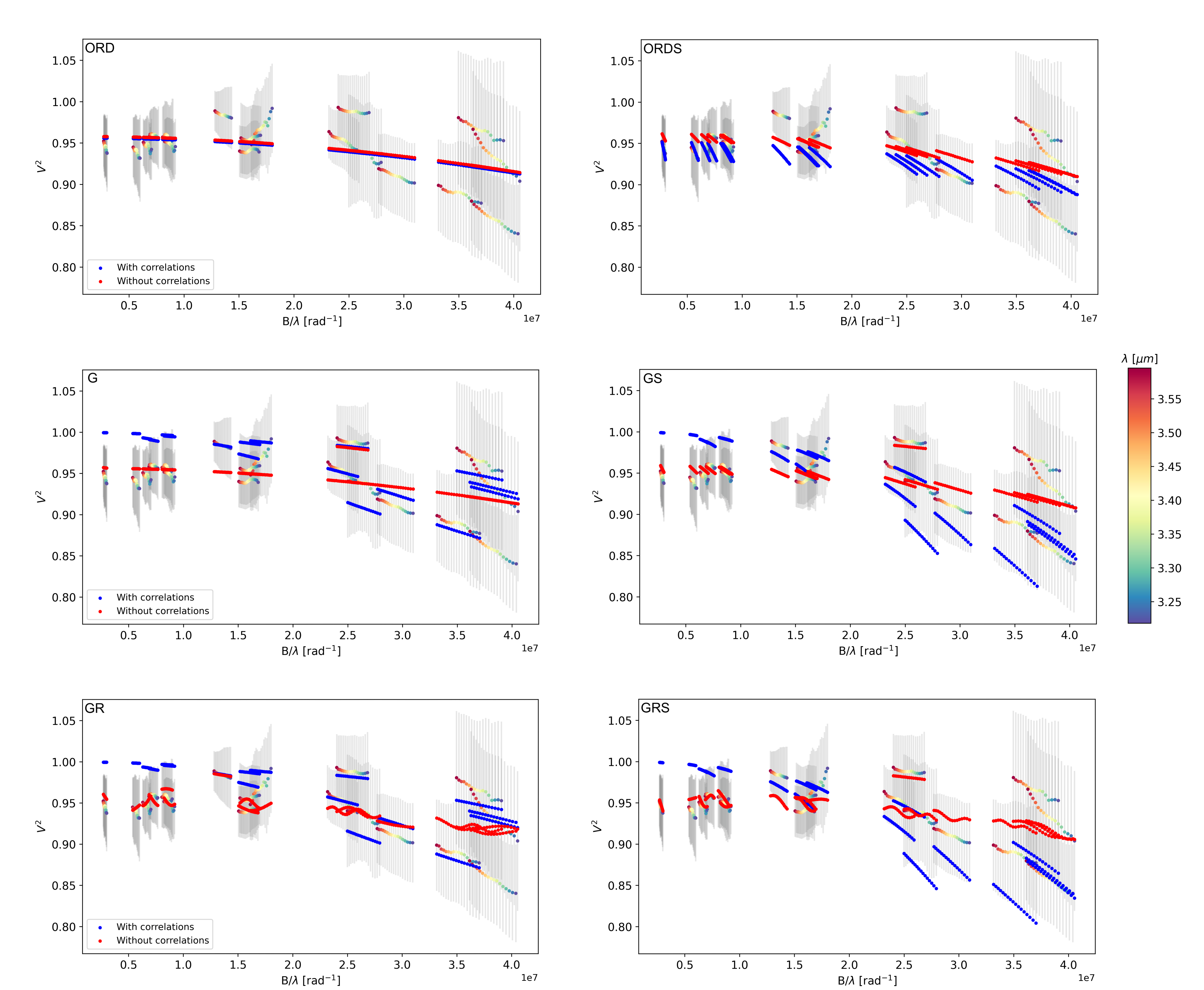}
    \caption{Best fits for the single-component models with (blue) and without (red) accounting for correlations. Left panels: Without spectral dependence. Right panels: With spectral dependence. The model keyword is indicated at the top left of the figure.}
    \label{fig:single_comp_fits}
\end{figure*}

\begin{figure*}[h!]
    \centering
    \includegraphics[width=1\textwidth]{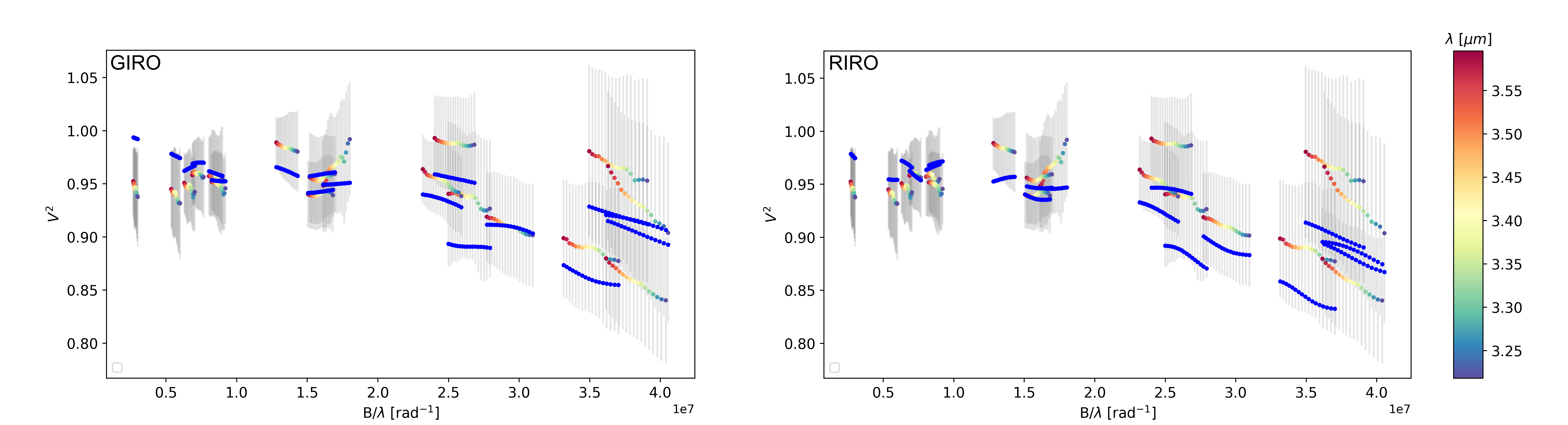}
    \caption{Best fits for the two-component models. The model keyword is indicated at the top left of the figure.}
    \label{fig:two_comp_fits}
\end{figure*}

Figures~\ref{fig:single_comp_fits} and \ref{fig:two_comp_fits} display the best-fit models to the \textit{L}-band VLTI/MATISSE visibilities obtained for $\beta$~Pictoris, for the single- and two-component models, respectively. \textbf{G} is the Gaussian model, \textbf{GR} is the Gaussian ring model, \textbf{GIRO} is the inner 2D Gaussian component with an outer ring component and \textbf{RIRO} is the two-Gaussian-ring model (inner and outer components). Model keywords ending in S denote models with a spectral dependence on the flux ratio.

\end{appendix}

\end{document}